\documentclass[longauth]{aa}  

\usepackage{graphicx}
\usepackage{txfonts}
\usepackage{hyperref}
\usepackage{xcolor}

\usepackage{natbib}

\DeclareUnicodeCharacter{2212}{-}

\newcommand{\cms}{cm/s}

\begin{document} 

   \title{Chromaticity of stellar activity in radial velocities}
    \subtitle{Anti-correlated families of lines on the M dwarf EV Lac with SPIRou and SOPHIE}

   \author{P. Larue\inst{\ref{inst.UGA}},
          X. Delfosse\inst{\ref{inst.UGA}},
          A. Carmona\inst{\ref{inst.UGA}},
          N. Meunier\inst{\ref{inst.UGA}},
          É. Artigau\inst{\ref{inst.UnivMontreal}\ref{inst.MontMegantic}},
          S. Bellotti\inst{\ref{inst.Leiden}},
          P. Charpentier\inst{\ref{inst.IRAP}},
          C. Moutou\inst{\ref{inst.IRAP}},
          J.-F. Donati\inst{\ref{inst.IRAP}},
          I. Boisse\inst{\ref{inst.LAM}},
          T. Forveille\inst{\ref{inst.UGA}},
            L. Arnold\inst{\ref{inst.CFHT}}
            V. Bourrier\inst{\ref{inst.Geneve}},  
            X. Bonfils\inst{\ref{inst.UGA}},  
            C. Cadieux\inst{\ref{inst.Trottier}},  
            A. Chomez\inst{\ref{inst.ParisMeudon}},  
            N. Cook\inst{\ref{inst.Trottier}},  
            P. Cortes Zuleta\inst{\ref{inst.StAndrews}},  
            P. Cristofari\inst{\ref{inst.Harvard}},  
            R. Diaz\inst{\ref{inst.Argentina}},  
            R. Doyon\inst{\ref{inst.Trottier}},  
            S. Grouffal\inst{\ref{inst.LAM}},  
            N. Hara\inst{\ref{inst.LAM}},  
            N. Heidari\inst{\ref{inst.Brazil}},  
            G. Hébrard\inst{\ref{inst.Paris}},  
            F. Kiefer\inst{\ref{inst.Paris}},  
            L. Mignon\inst{\ref{inst.UGA}},  
            A. Maurel\inst{\ref{inst.Paris}},  
            J. Morin\inst{\ref{inst.Montpellier}},  
            A. Petit\inst{\ref{inst.Nice}},
            P. Petit\inst{\ref{inst.IRAP}},    
            A. Santerne\inst{\ref{inst.LAM}},  
            N. Santos\inst{\ref{inst.Porto1}\ref{inst.Porto2}},  
            D. Segransan\inst{\ref{inst.Geneve}},  
            J. Serrano Bell\inst{\ref{inst.Argentina}},  
            H. G. Vivien\inst{\ref{inst.LAM}}
}

    \institute{Univ. Grenoble Alpes, CNRS, IPAG, 38000 Grenoble, France \label{inst.UGA}
    \and
    Université de Montréal, Département de Physique, IREX, Montréal, QC H3C 3J7, Canada \label{inst.UnivMontreal}
    \and
    Observatoire du Mont-Mégantic, Université de Montréal, Montréal, QC H3C 3J7, Canada \label{inst.MontMegantic}
    \and
    Leiden Observatory, Leiden University, PO Box 9513, 2300 RA Leiden, The Netherlands \label{inst.Leiden}
    \and
    Univ. de Toulouse, CNRS, IRAP, 14 av. Belin, 31400 Toulouse, France \label{inst.IRAP}
    \and
    Aix Marseille Univ., CNRS, CNES, LAM, Marseille, France \label{inst.LAM}
    \and
    Canada France Hawaii Telescope Corporation (CFHT), UAR2208 CNRS-INSU, 65-1238 Mamalahoa Hwy, Kamuela 96743 HI, USA \label{inst.CFHT}
    \and
    Observatoire Astronomique de l’Université de Genève, Chemin Pegasi 51b, 1290 Versoix, Switzerland \label{inst.Geneve}
    \and
    Institut Trottier de recherche sur les exoplanètes, Université de Montréal, 1375 Ave Thérèse-Lavoie-Roux, Montréal, QC, H2V 0B3, Canada \label{inst.Trottier}
    \and
    LIRA, Observatoire de Paris, Université PSL, CNRS, Université Paris Cité, Sorbonne Université, 5 place Jules Janssen, 92195 Meudon, France \label{inst.ParisMeudon}
    \and
    SUPA School of Physics and Astronomy, University of St Andrews, North Haugh, St Andrews KY16 9SS, UK \label{inst.StAndrews}
    \and
    Center for Astrophysics | Harvard \& Smithsonian, 60 Garden Street, Cambridge, MA 02138, USA \label{inst.Harvard}
    \and
    International Center for Advanced Studies and ICIFI (CONICET), ECyT-UNSAM, Campus Miguelete, 25 de Mayo y Francia, (1650), Buenos Aires, Argentina \label{inst.Argentina}
    \and
    Laboratório Nacional de Astrofísica, Rua Estados Unidos 154, 37504-364, Itajubá-MG, Brazil \label{inst.Brazil}
    \and
    Institut d’Astrophysique de Paris, CNRS, UMR 7095, Sorbonne Université, 98 bis bd Arago, 75014 Paris, France \label{inst.Paris}
    \and
    Laboratoire Univers et Particules de Montpellier, Université de Montpellier, CNRS, 34095 Montpellier, France \label{inst.Montpellier}
    \and
    Université Côte d’Azur, Observatoire de la Côte d’Azur, CNRS, Laboratoire Lagrange, France \label{inst.Nice}
    \and
    Instituto de Astrofísica e Ciências do Espaço, CAUP, Universidade do Porto, Rua das Estrelas, 4150-762 Porto, Portugal \label{inst.Porto1}
    \and
    Departamento de Física e Astronomia, Faculdade de Ciências, Universidade do Porto, Rua do Campo Alegre, 4169-007 Porto, Portugal \label{inst.Porto2}
    }

   \date{Received 1 April 2025 / Accepted 16 July 2025}
 
  \abstract
   {In the search for exoplanets using radial velocities (RV), stellar activity has become one of the main limiting factors for detectability. Fortunately, activity-induced RV signals are wavelength-dependent or chromatic, unlike planetary signals. This study exploits the broad spectral coverage provided by the combined use of SOPHIE and SPIRou velocimeters to investigate the chromatic nature of the activity signal of the highly active M dwarf EV Lac.}
   {We aim to understand the origin of the strong wavelength dependence (chromaticity) observed in the RV signal of EV Lac by selecting spectral lines based on physical properties. In particular, we explore the impact of starspots by defining the contrast effect at the level of individual lines. The Zeeman effect is also considered in this study.}
   {SPIRou and SOPHIE spectra were reduced using the line-by-line (LBL) method. We performed custom RV calculations, using groups of spectral lines selected for their sensitivity to either the spot-to-photosphere contrast or the Zeeman effect. The sensitivity of each line to the spot is defined using a two-temperature model based on PHOENIX spectra, while Landé factors were used to quantify Zeeman sensitivity.}
   {We find that the spectral lines are distributed in two distinct families of contrasts,  producing anti-correlated RV signals. This leads to a partial cancellation of the total RV signal, especially at longer wavelengths and provides a natural explanation for the strong chromaticity observed in EV Lac. This sign-reversal effect is demonstrated here, for the first time, on empirical data. Building on this discovery, we propose a new approach to constraining spot temperatures and to mitigating stellar activity. This will open up promising avenues for improving activity corrections and enhancing the detection of exoplanets around active M dwarfs.}
  {}

   \titlerunning{Anti-correlated families of lines on the M dwarf EV Lac}
   \authorrunning{P. Larue}
   \maketitle

\section{Introduction}\label{Sect.Intro}
  
Searching for planets around M dwarfs using the radial velocity (RV) method is highly effective, in particular (but not exclusively) to discover low-mass planets with moderate temperatures \citep[e.g.,][]{Bonfils_2013_occurence}.
This is true not only because of the lower star-planet mass ratio, but also because the habitable zone is closer to the central star than for systems orbiting solar-type stars, so these planets have shorter orbital periods, facilitating their discovery.\\ \indent
Moreover, 61\% of the objects in the sample of stars and brown dwarfs within 10\,pc of the Sun are M dwarfs \citep{Reyle_2021_Gaia10pc}. Statistical studies, obtained with transiting planets (\cite{Dressing_2015_occurence-small-planets-M, Hsu_2020_statistics_transit, Ment_2023_occurence_transit} or from RV surveys \citet{Bonfils_2013_occurence, Sabotta_2021_occurence, Mignon_occurence}), show that these stars should each host around 1.6 planets with masses between 1 and 31 Earth masses and periods of less than 100 days in average. Nonetheless, among the closest 100 M dwarfs \citep{Reyle_2021_Gaia10pc} only 29 are known to host planetary systems\footnote{As of April 1, 2025, by cross-referencing the \citet{Reyle_2021_Gaia10pc} Vizier catalog with the NASA Exoplanet Archive: \href{https://exoplanetarchive.ipac.caltech.edu}{exoplanetarchive.ipac.caltech.edu}}, despite the continuous improvement of the sensibility of spectrographs, and the advent for Near Infra-Red (NIR) spectroscopy in recent decades (e.g., HPF \citep{Mahadevan_2019_HPF}, CARMENES-NIR \citep{Bauer_2020_CARMENES}, IRD \citep{Kotani_2018_IRD}, SPIRou \citep{Donati_2020_SPIRou}, and NIRPS, \citealt{Bouchy_2017_NIRPS}).\\ \indent
Thus, many low-mass planets remain to be discovered and one of the main reasons for the lack of detection seems to be stellar activity \citep{Barnes_2014_RV-precision-activity, John_2023_subms_activity}. M dwarfs are particularly affected, since a large fraction of them show high levels of activity \citep[e.g.][]{Delfosse1998, Reiners_activity_2010, Reiners_activity_2012, Mascareno_activity_2016}, especially the less massive ones, which have a higher proportion of fast rotators due to their longer spin-down timescales. This poses significant challenges to the detection of exoplanets around these stars and introduces biases into statistical studies. An additional complication arises from the fact that M dwarfs often have rotation periods similar to the orbital periods of planets in their habitable zones \citep{Newton_2016_M-HZ-detectability}.\\ \indent
The different activity processes associated with the stellar magnetic field can be manifested in various forms such as spots, plages, inhibited convection, or the Zeeman effect. Occurring at various spatial and temporal scales, they have a direct impact on the RV measurements \citep{Saar_1997_activityRV, Boisse2011}. This means that they can either introduce noise \citep{Meunier_2010a_spotRV, Meunier_2021_flowsRV} or even mimic false planetary signals, as in the case of AD Leo, where activity led to a spurious detection in optical RVs, which was subsequently invalidated by NIR observations \citep[see][]{Carmona_ADLeo}.\\  \indent
Depending on the spectral type and the level of activity, the resulting RV jitter can range from tens of \cms \citep{Meunier_2021_lecture} to hundreds of m/s \citep{Klein2022}. Disentangling this activity-induced signal from the Keplerian signal of a planet can be complex, yet it is crucial for detecting planets that are still hidden, particularly in the solar neighborhood.\\ \indent
In this paper, we  analyze both the Zeeman effect from highly magnetic regions and the impact of dark spots on the surface of a $0.32\,M_\odot$ active M dwarf. As these features rotate with the star, they result in a Doppler shift variation modulated by the star’s rotational period.\\  \indent
The effect of dark spots is wavelength dependent, with the spot-induced jitter being less pronounced in the NIR compared to the optical range \citep[e.g.,][]{Huelamo2008, Figueira2010, Mahmud2011, Zechmeister_2018_CRX, Carmona_ADLeo}. This wavelength dependence, known as the chromaticity of the activity signal, can arise in various forms and from different stellar processes. More specifically, even lines of similar wavelength can respond differently to a given stellar feature. That is why previous studies have looked into the inhomogeneous influence of activity on different spectral lines \citep[e.g.][]{Lopez-Gallifa_2021_EVLac-sensible-lines-poster, Bellotti_2022_line-selection-activity}. The work presented here draws on this approach.\\  \indent
The aim of this study is to further our understanding of the physical processes driving RV variations across different wavelengths in the case of an active M Dwarf. An ideal target for this investigation is EV Lac (GJ\,873), a source we have quasi-simultaneously monitored over three years with the SPIRou NIR spectrometer located at the Canada France Hawai Telescope (CFHT, \citealt{Donati_2020_SPIRou}) and the SOPHIE optical spectrometer located at the Observatoire de Haute Provence (OHP, France, \citealt{Bouchy_2013_SOPHIE+}). The use of these two velocimeters enables us to study stellar activity signals over a broad wavelength range, providing significant leverage for investigating global chromatic effects. With the aim of linking individual spectral line variations to distinct stellar processes, we calculated the RVs using a line-by-line approach \citep[LBL,][]{Artigau_LBL2022}, inspired by \citet{Dumusque_2018_LBL_pRVs_I} and \citet{Cretignier_2020_LBL_pRVs_II}. Although this approach is similar to that of \citet{Bellotti_2022_line-selection-activity}, here the selection of lines based on physical criteria was taken a step further. First, the LBL method provides more lines for us to work with, enabling more subtle line selections while maintaining precise RV measurements. Additionally, we refined the selection of spectral lines based on physical properties, particularly by considering the variation in absorption lines between the photosphere and spots. Ultimately, given the goal is not only to obtain activity-sensitive and non-sensitive samples of lines, but more specifically to link line variations to distinct stellar processes, this approach could provide new insights into the stellar physics of active M dwarfs. In particular, it could help to constrain the temperature and Zeeman effect of starspots, as well as their respective impact on spot spectra. This could lead to new ways of reducing the impact of stellar activity through physics-driven line selections.\\  \indent
The layout of the paper is as follows. The target EV Lac is presented in Section \ref{Sect.EVLac}. Particular emphasis is placed on information relevant to the activity, such as the star's rotation period, magnetic field, and spot temperature. Section \ref{Sect.Obs} describes the observations obtained with the two velocimeters SPIRou and SOPHIE, as well as how the RVs are extracted. Section \ref{Sect.Methodology} details the line selection methodology. The chromaticity of the RV signal is shown in Section \ref{Sect.Chromaticity}, and explained in Section \ref{Sect.Contrast}. Section \ref{Sect.Zeeman} explores the chromaticity of the Zeeman effect, while Section \ref{Sect.Correction} explores a new way of measuring the spot temperature, and  an innovative activity mitigation technique. Finally, Section \ref{Sect.Conclusion} summarizes the main conclusions and discusses potential avenues for future studies.

\section{EV Lac}\label{Sect.EVLac}

\subsection{Stellar properties}\label{subSect.EVLac1}

At a distance of 5.049 ± 0.001 pc \citep{GAIA_coord}, in the Lacertae constellation, the very low mass star EV Lacertae (EV Lac, GJ 873) is one of the 60 nearest stellar systems. Its stellar parameters are listed in Table \ref{Table_Stellar_params}.
It is one of the most studied stars due to its high flare activity \citep[e.g.,][]{Paudel2021}. Based on spectral indices, \citet{Shkolnik2009} estimated its age to be between 25 and 300 Myr. This is broadly consistent with the classification of EV Lac as a member of the Ursa Major Moving Group \citep{Klutsch2014, Cortes_Contreras2024}, whose age is estimated to be around 300–400 Myr \citep{Soderblom1993, Jones2015}.

\begin{table}[h]
    \caption[]{EV Lac stellar properties}
    \label{Table_Stellar_params}
    \begin{tabular}{p{0.29\linewidth}p{0.25\linewidth}p{0.18\linewidth}p{0.1\linewidth}}
        \hline
        \hline
        \noalign{\smallskip}
        \textbf{Parameter}      &  \textbf{Value}   & \textbf{Units} &   \textbf{Ref.} \\
        \noalign{\smallskip}
        \hline
        \hline
        \noalign{\smallskip}
        \multicolumn{4}{c}{Astrometric properties}\\
        \hline
        \noalign{\smallskip}
        Names           &   GJ873, EV Lac  &  &  \\
        $\alpha$ (J2000)  &   22:46:49.73         & deg & 1 \\
        $\delta$ (J2000)  &   +44:20:02.4         & deg & 1 \\
        $\mu_{\alpha}$ & -706.1$\pm$0.1 & mas.yr$^{-1}$ & 1 \\
        $\mu_{\delta}$ & -458.8$\pm$0.1 & mas.yr$^{-1}$ & 1 \\
        parallax & 198.01$\pm$0.04 & mas & 1 \\
        distance          &   $5.049 \pm 0.001$ & pc & 1 \\
        \hline
        \noalign{\smallskip}
        \multicolumn{4}{c}{Physical parameters}\\
        \hline
        \noalign{\smallskip}
        Spectral type       &   M3.5 V      &   & 2 \\
        Mass & $0.32 \pm 0.02$ & $M_\odot$  & 3 \\
        Radius    &   $0.31 \pm 0.02$ & $R_\odot$   & 3 \\
        log {\it g} & 4.87$\pm$0.05 & $\log$(cm.s$^{-2}$) & 3 \\
        Age         & 25 - 300      & Myr               &    4   \\
        $P_{\text{rot}}$   &   $4.359 \pm 0.002$ & d  & 5 \\
        v sin(i) &   $4.1$ & km.s$^{-1}$ & 6 \\
        i       &   $66 \pm 28$ & degree  & 7 \\
        $T_{\text{eff}}$  & $3340 \pm 31$ & K   & 3 \\
        $T_{\text{spot}}$ (theoretical)  & $2733 \pm 542$ & K   & 8 \\
        $T_{\text{spot}}$ (photometry)  & $3268$ & K   & 9 \\
        Small scale $<B>$ & $4.53 \pm 0.07$ & kG & 3 \\
        Large scale $B_{l}$ & $300 - 500$ & G & 10, 11 \\
        Fe/H & -0.01 $\pm$ 0.15 & dex & 12 \\
        EW H$\alpha$ & -4.54$\pm$0.04 & \AA & 13 \\
        EW Ca II K & 14.86 & \AA & 14 \\
        log $L_{H \alpha}$/$L_{bol}$ & -3.76 & & 13 \\
        log $L_{X}$/$L_{bol}$ & -3.33 & & 10 \\
        log $R'_{HK}$ & -4.24$\pm$0.11 & & 15 \\
        \hline
        \noalign{\smallskip}
    \end{tabular}\\ 
    \tablebib{
    (1) \citet{GAIA_coord}, 
    (2) \citet{Reid_1995_Spectral-type}, 
    (3) \citet{Cristofari_2023_parameters}, 
    (4) \citet{Shkolnik2009}, 
    (5) \citet{Paudel_2021_EV-Lac}, 
    (6) \citet{Reiners_2022_vsini}, 
    (7) \citet{Reiners_2018_CARMENES-vsini}, 
    (8) \citet{herbst_2021_starspots}, 
    (9) \citet{Ikuta_2023_spots}, 
    (10) \citet{Morin_2008_M-B-fields}, 
    (11) \citet{Bellotti_2024_EVLac-Bfield}, 
    (12) \citet{Rojas-Ayala_2012_metallicity}, 
    (13) \citet{Newton_2017_Halpha}, 
    (14) \citet{Youngblood_2017_Ca}, 
    (15) \citet{Melbourne_2020_RHK}
    }\\ 
\end{table}

\subsection{Stellar activity}\label{subSect.EVLac2}

EV Lac is a very active star, one of the most active M dwarfs, with $log(R'_{HK})$ values as high as $-4.24\pm0.11$ \citep{Melbourne_2020_RHK}, corresponding to saturation values for the Ca II H and K doublet emission for $P_{rot} < 10$ days \citep{Astudillo-Defru2017}. Its activity is characterized by numerous flares \citep[e.g.,][]{Favata_2000_EVLac-Xray-flare, Honda_2018_optical-flare, Paudel_2021_EV-Lac, Inoue_2024_EVLac-flares, Chen_2022_EVLac-Flares}, the presence of spots on its surface \citep[e.g.,][]{Ikuta_2023_spots}, and one of the highest average surface magnetic fluxes among M dwarfs \citep[e.g.,][]{Morin_2008_M-B-fields, Shulyak_2019_CARMENES-Bfields, Cristofari_2023_parameters, Bellotti_2024_EVLac-Bfield}.\\  \indent
Examining its magnetic field geometry is particularly relevant, as highly magnetized regions directly contribute to the formation of spots, which are key to our study. The large-scale magnetic field of EV Lac derived from Stokes V profiles is non-axisymmetric, mainly poloïdal with an average strength of 300 G from 2020 to 2021 \citep{Bellotti_2024_EVLac-Bfield}. Although it has evolved over time, Zeeman doppler imaging (ZDI) reconstructions of the magnetic field have revealed a non-axisymmetric dipole, which is likely responsible for the formation of two spots at opposite longitudes
\citep{Morin_2008_M-B-fields, Shulyak_2019_CARMENES-Bfields, Bellotti_2024_EVLac-Bfield}.\\  \indent 
The star is expected to be in solid body rotation \citep{Morin_2008_M-B-fields} with an inclination of $66 \pm 28$ degrees and a vsin\textit{i} of 4.1 km/s \citep{Reiners_2022_vsini}. Studies give values of $P_{rot} = 4.3615 \pm 0.0001 $ days \citep{Morin_2008_M-B-fields}, $P_{rot} = 4.38 \pm 0.03 $ days \citep{Reiners_2018_CARMENES-vsini}, $P_{rot} = 4.379$ days \citep{Diez-Alonso_2019_Prot}, $P_{rot} = 4.3592$ days \citep{Paudel_2021_EV-Lac}, and $P_{rot} = 4.36 \pm 0.05 $ days in \citet{Bellotti_2022_line-selection-activity}. The small differences in rotation periods can be physically justified, for example, by differential rotation or variations in the latitude of active regions. Here, we adopted the latest value of 4.36 days, which is in very good agreement with the activity signal observed in our SOPHIE RV measurements (see Fig. \ref{Fig1_SOPHIE-SPIRou_RVs}).\\  \indent
Estimating the spot temperature is particularly complex, but is of great importance to our study. \citet{herbst_2021_starspots} has updated the empirical relation of \citet{Berdyugina_2005}, which estimates the temperature difference between the photosphere and the spots for several stars and derives a polynomial fit as a function of the stellar effective temperature. However, they used a sample of 40 stars, of which only 4 are M dwarfs, and the resulting uncertainties are high: using their empirical law for EV Lac gives a spot temperature of $2764 \pm 517$\,K, so a temperature difference between the spot and the photosphere of $\sim 575 \pm 520$\,K. Another study from \citet{Ikuta_2023_spots} used photometric measurements from TESS to estimate the filling factor of the spots and deduced an estimated spot temperature of 3268\,K, so a temperature difference between the spot and the photosphere of $\sim 72$\,K. However, these measurements must be interpreted with caution, as the problem is degenerate between the spot temperature and its area. These differences indicate that the spot temperature is poorly constrained. As a starting point, we adopt an intermediate value of 3000\,K, before revisiting this estimate in Section \ref{Sect.Correction} of this article.

\section{Spectroscopic observations}\label{Sect.Obs}

\begin{figure}[]
    \centering
    \includegraphics[width=1\linewidth]{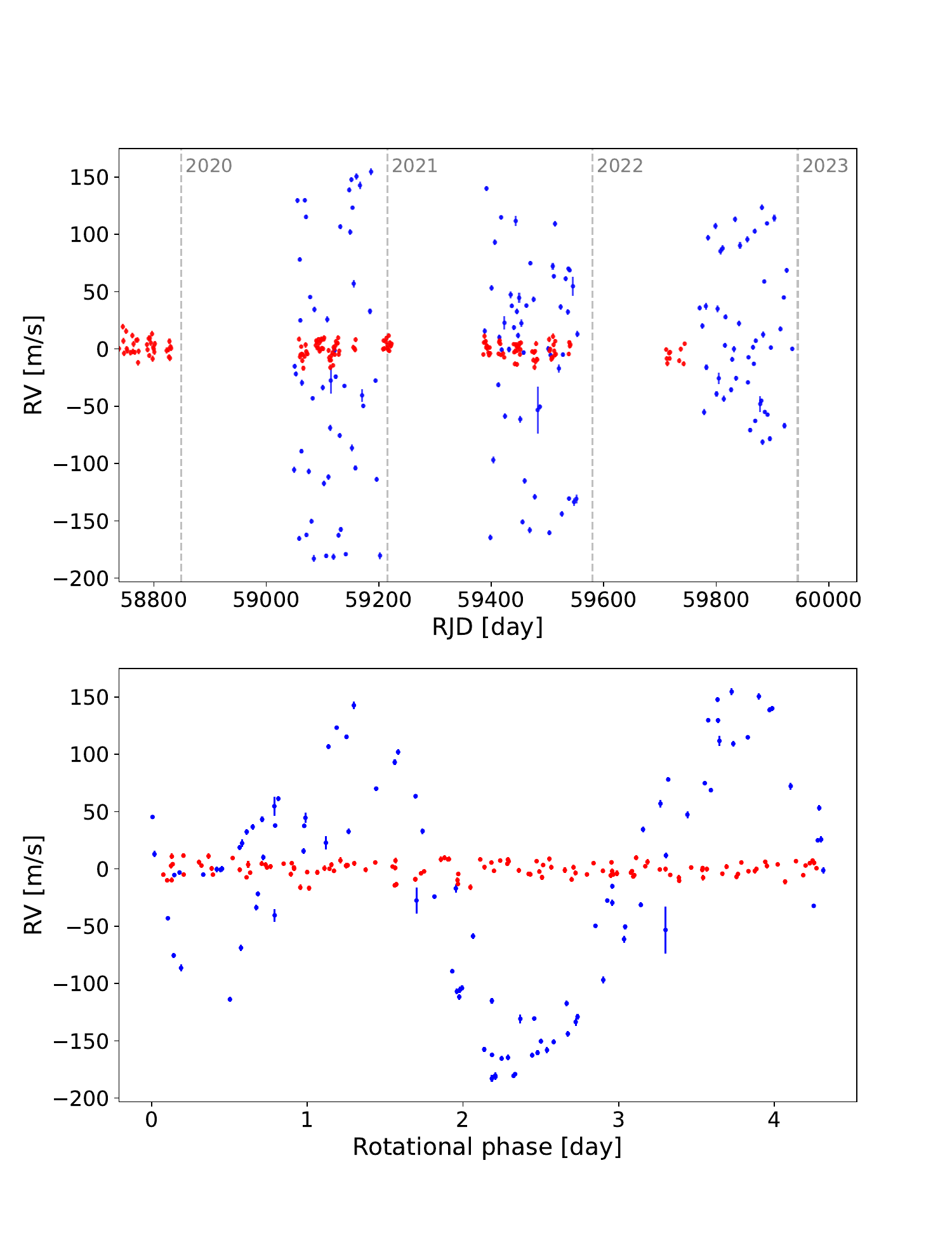}
    \caption{SOPHIE measurements of EV Lac, reduced by LBL shown in blue. SPIRou measurements, reduced by LBL shown in red. \textbf{Top panel:} Full dataset of SOPHIE and SPIRou RVs. \textbf{Bottom panel:} EV Lac RVs phase-folded on the rotation period, for SOPHIE (RMS = \hbox{95.7\,m/s}, mean error = \hbox{2.8\,m/s}) and SPIRou (RMS = \hbox{6.3\,m/s}, mean error = \hbox{2.1\,m/s}) concomitant seasons only (2020 and 2021). Note: these data were reduced using the entire LBL pipeline of \citet{Artigau_LBL2022}, including the statistical line rejections.}
    \label{Fig1_SOPHIE-SPIRou_RVs}
\end{figure}

We used spectroscopic observations of EV Lac spanning three years, with a total of 317 nights of observations obtained with two high-resolution velocimeters: SOPHIE in the optical \citep{Bouchy_2006_SOPHIE, Perruchot_2008_SOPHIE, Bouchy_2013_SOPHIE+} and SPIRou in the NIR \citep{Donati_2020_SPIRou}. The RV dataset is presented in Fig. \ref{Fig1_SOPHIE-SPIRou_RVs}, which shows that the SOPHIE and SPIRou measurements are quasi-simultaneous only in 2020 and 2021. Since the aim of this paper is to investigate the wavelength dependence of the effect of stellar activity on RV and considering that activity processes vary over time, it is essential to ensure that both instruments are measuring this effect simultaneously. For this reason, the rest of this paper  only includes data obtained by SOPHIE and SPIRou during the same periods, namely, the years 2020 and 2021.

\subsection{High-resolution spectroscopy with SPIRou}\label{subSect.Obs3}

SPIRou (Spectro-Polarimetre InfraRouge) is a high-resolution spectro-polarimeter and velocimeter mounted on the Canada-France-Hawaii Telescope (CFHT, \citealt{Donati_2020_SPIRou}). It is fiber-fed from the Cassegrain focus and is built to obtain very high RV precision, on the order of meters per second over several years. SPIRou is designed as an echelle spectrograph that allows to observe a reference spectrum simultaneously to the object of interest. It covers the NIR spectral domain from 950\,nm to 2350\,nm (corresponding to the Y, J, H, and K bands) across 50 spectral orders, for a spectral resolution of $R\approx70\,000$. It also includes a polarimeter to analyze the polarization of spectral lines of the incoming stellar light. As our observations of EV Lac are obtained in this polarimetric mode, each SPIRou observation is therefore a sequence of four subexposures, each one with specific azimuths of the twin quarter-wave Fresnel rhombs.\\  \indent
EV Lac measurements were taken between September 2019 and June 2022, for a total of 169 nights of observation,  reaching a mean signal-to-noise ratio (S/N) per pixel of 131 in the 44\textsuperscript{th} grating diffraction order (H band, at 1\,746\,nm). These observations were conducted as part of the SPIRou Legacy Survey Planet Search program (SLS-PS) \citep{Moutou_2023_SPIRou-SLS}.\\ \indent
The spectra were reduced using the nominal APERO pipeline version 0.7.286 \citep{Cook_2022_APERO-SPIRou}, and the RVs were computed in the telluric corrected spectra with the LBL algorithm version 0.63.002 for precision RV measurements \citep{Artigau_LBL2022}. This method efficiently handles outlying spectral information by leveraging all available observations to construct a high-precision median template spectrum of the star, where the spectral lines are defined as the regions between two consecutive local maxima. Then, the local RV and associated error were computed for each line, using \citet{Bouchy_2001_equations} equations. At this stage, it is possible to statistically reject lines that behave as outliers. For that purpose, LBL uses a simple mixture model to estimate the mean velocity, where there are two distributions, one for valid lines that follow a normal distribution and another broader distribution corresponding to outliers.

\subsection{High-resolution spectroscopy with SOPHIE}\label{subSect.Obs1}

SOPHIE is a high-resolution fiber-fed spectrograph and velocimeter mounted on the 1.93m telescope at the Observatoire de Haute Provence (OHP, France). Its first light was achieved in 2006 \citep{Bouchy_2006_SOPHIE}, but it received a major update in 2011 with the implementation of an octagonal-section fiber in the fiber link \citep{Perruchot_2011_SOPHIE-fiber}, enabling SOPHIE+ to reach the meter-per-second stability \citep{Bouchy_2013_SOPHIE+}.\\  \indent
Two observation modes are available, offering different throughput and spectral resolution via two independent fiber feeds. The measurements presented in this paper were obtained through the high resolution (HR) mode, used with a simultaneous Fabry-Perot calibration lamp, achieving a spectral resolution of $R\approx75000$. This resolution is estimated at 550\,nm, while the spectrograph covers a wavelength range from 387.2\,nm to 694.3\,nm, across 38 spectral orders.\\  \indent
EV Lac measurements were taken between August 2020 and December 2022, for a total of 148 nights of observation, and reach a mean S/N per pixel of 87 in the 35\textsuperscript{th} order. Observations were conducted as part of the sub-program three (SP3) of a very large RV survey, initiated in 2006 \citep{Bouchy_2009_DRS}, which targets M dwarfs in the solar neighborhood to detect new planets \citep{Hobson_2018_NAIRA, Hobson_2019, Diaz_2019_Gl411, Cortes_2024_gl725} or to characterize their activity \citep[e.g.,][]{Carmona_ADLeo, Cortes_2023_gl205}.\\  \indent
The spectra were reduced with the SOPHIE DRS \citep{Bouchy_2009_DRS}, which performs spectral orders localization, optimal order extraction, cosmic-ray rejection, and wavelength calibration. It also applies the cross-correlation function (CCF) technique to derive the RVs; in this case, by cross-correlating an empirical M4 mask to the observed spectra. From this, the CCF is then computed standard activity indicators such as the bissector (BIS) or the full width at half maximum (FWHM). We also computed the activity indices based on the $H_\alpha$ line and the Ca II H\&K lines (S index).  To be consistent with the RVs obtained with SPIRou, we computed them with a LBL pipeline on the SOPHIE data as described below.

\subsection{The SOPHIE-LBL pipeline}\label{subSect.Obs2}

To adapt the LBL pipeline for use with SOPHIE data, it is essential to ensure the input consists of high-quality, homogeneous observations. The preprocessing stage of the pipeline begins by filtering the available data, retaining only observations acquired in HR mode. The RV derived from the CCF and its FWHM, provided by the SOPHIE DRS \citep{Bouchy_2009_DRS}, are used to perform an additional 4$\sigma$ rejection of outliers. Once these spectra have been collected, they are corrected for the charge transfer inefficiency (CTI) effect, following the method described in \citet{Bouchy_2009_CTI}. Then,  the LBL routine is applied to calculate the RVs and proxies of activity indicators \citep[see][]{Artigau_LBL2022} for each observation, but also for each line of a given observation.\\  \indent
Lastly, SOPHIE experiences a well-monitored RV zero-point variation over typical timescales ranging from a few days to several months, with an amplitude that can reach up to \hbox{10\,m/s}. To correct it, nightly observations of a list of RV standard stars (i.e., bright stars known to display no RV variations) are performed, using HD\,185144, HD\,9407, HD\,89269A and Gl\,514. This correction follows the procedure described in \citet{Courcol_2015_SOPHIE-master, Heidari_2022_SOPHIE-ZP}. The standard measurements are averaged with a sliding window over 30 measurements, with breaks allowed on dates where an instrumental change has been made \citep[see][p75 for a list of the dates]{Heidari_2022_PHD}.\\  \indent
This results in a root mean square (RMS) of the RV standard stars after the master correction (namely a proxy of the average RV uncertainty) of \hbox{2.09\,m/s}, which is rather constant over the master time span (in the range of mjd = 55856 to 60151 at the time of this study).

\section{Line selection methodology}\label{Sect.Methodology}

The LBL framework generates a file for each observation, which contains the per-line information. These files  not only contain the per-line RVs, but also the projection of the second and third derivatives of the line profile, which are proxies of  the variation of the FWHM of the line and its bisector \citep{Artigau_LBL2022}, respectively. In our analysis, we  used these files directly instead of the final output of LBL, which applies a statistical rejection process that we aimed to bypass in order to prioritize line selection based on physical criteria. However, we implemented one systematic rejection: for each line, we computed its RV error averaged over all observations. Some lines are clear outliers from this mean error, either due to excessive noise or an unrealistically low RV error\footnote{These outliers are due to the current version of LBL, which can be improved in the futur.}, which could bias the overall weighted mean RV. These outliers were identified and manually removed.\\  \indent
We defined the skeleton of our selection process as the list of remaining lines and their spectral shapes, extracted from the stellar median spectrum. This served as a foundation, enabling us to assign each line the physical quantities that trace the phenomena we aimed to investigate. In the case of the present study two physical effects are explored: the impact on the RVs of spectrum contrast between the spots and the photosphere, as well as the impact of the Zeeman effect. The estimators used to study these two effects are  described in Sections \ref{subSect.Contrast2} and \ref{Sect.Zeeman}, respectively. Ultimately, we selected the lines based on any physical condition to obtain a new physics-based list of lines, then used to compute the resulting RV of each epoch using the finite mixture model used in LBL.

\section{EV Lac's chromaticity}\label{Sect.Chromaticity}

\subsection{Activity signal in SOPHIE and SPIRou datasets}\label{subSect.Chromaticity1}

\begin{figure}[]
    \centering
    \includegraphics[width=0.9\linewidth]{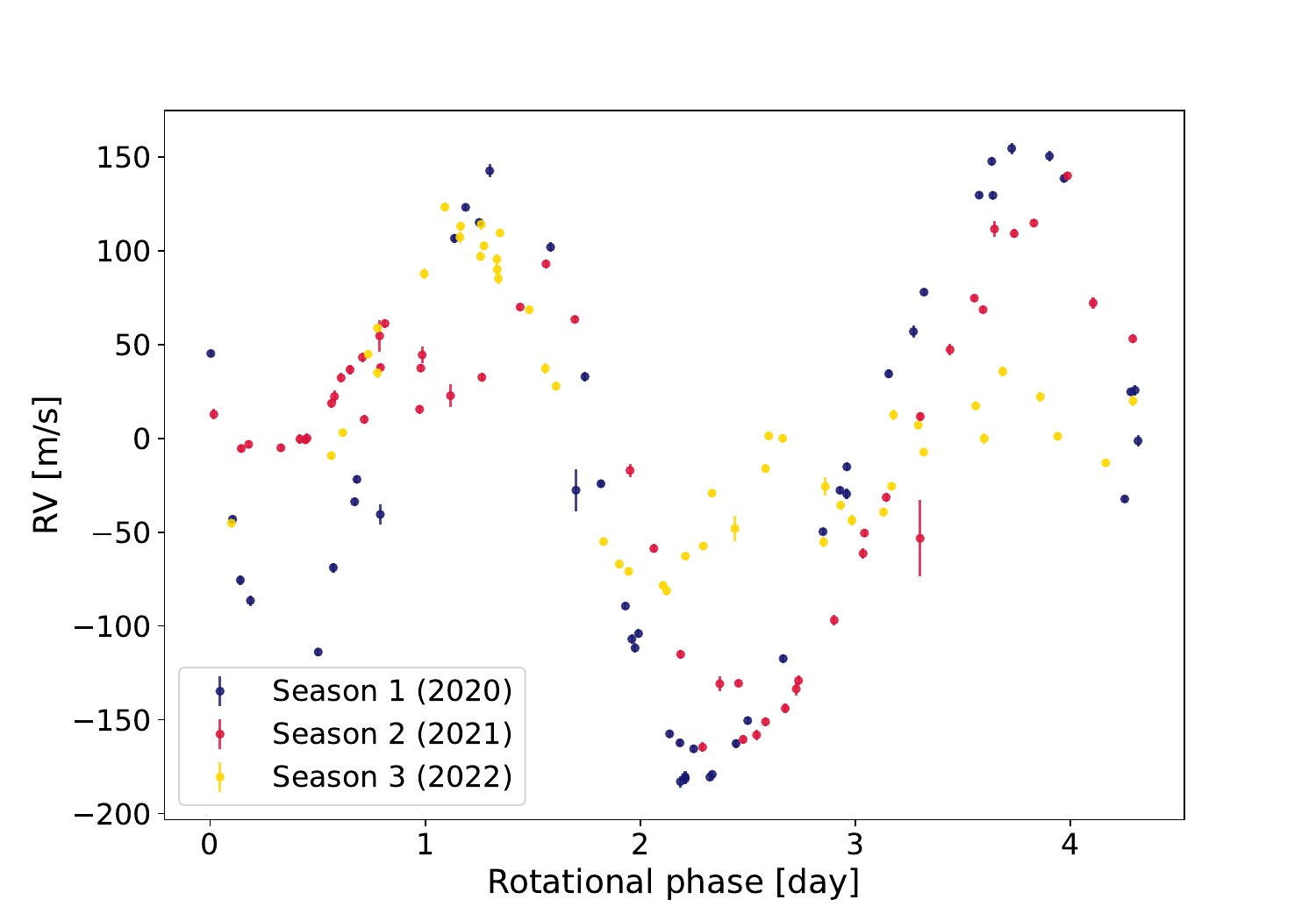}
    \caption{SOPHIE phase-folded RVs over the three seasons. Measurements taken in 2020 (RMS = \hbox{107.7\,m/s}) are shown in blue. Measurements taken in 2021 (RMS = \hbox{81.6\,m/s}) in red. \textbf{} Measurements taken in 2022 (RMS = \hbox{59.8\,m/s}) in yellow.}
    \label{Fig3_SOPHIE_seasons}
\end{figure}

The first analysis consisted of selecting the wavelength ranges corresponding to the two velocimeters to compare the optical and NIR. Looking at the two seasons where the star was observed both by SOPHIE and SPIRou (2020 and 2021), we observed (as illustrated in the top panel of Fig. \ref{Fig1_SOPHIE-SPIRou_RVs}) a strong chromaticity. In the optical, there is a signal with a RMS of \hbox{95.7\,m/s}, while in NIR, the RMS of \hbox{6.3\,m/s} is slightly higher than the uncertainty of the average measurement of \hbox{2.1\,m/s}. This yields a reduction in the RMS by a factor of 15 from optical to NIR. The SOPHIE periodogram (see Fig. \ref{A.A:SOPHIE_perio}) shows only signals at the rotation period and its harmonics, half the rotation period being the strongest one. In contrast, the SPIRou periodogram (see Fig. \ref{A.A:SPIRou_perio}) does not show a peak in the rotation period, nor in its harmonics. When the two datasets from these two seasons are phase-folded at the rotation period (see bottom panel of Fig. \ref{Fig1_SOPHIE-SPIRou_RVs}), a clear signal appears in the SOPHIE RVs, whereas the SPIRou RVs show a more stochastic signal without RV modulation.\\ \indent
The strong chromaticity of the RV signal confirms its activity origin (as in previous studies on similar targets; see, e.g., \cite{Carmona_ADLeo}). In this case, it is attributed to the presence of two dominant spots shifted in phase by half a cycle (as discussed in Section \ref{subSect.EVLac2}). However, from a theoretical point of view, it is difficult to predict such a factor of 15 in the amplitude reduction with wavelength if the chromaticity is only due to the luminosity difference between the photosphere and the spot.
For instance, \citet{Reiners_2010_spots} simulated the impact of spots at different temperatures on RV chromaticity. Their model generated a $500\times500$ grid of PHOENIX spectra with two temperatures, representing the photosphere and circular spots at $0^{\circ}$ longitude and $+30^{\circ}$ latitude, covering 1\% of the projected stellar surface. They accounted for rotation-induced projection effects, assuming a stellar inclination of $90^{\circ}$ and no limb darkening. RVs were computed across different temperatures and rotational velocities in 50\,nm-wide bins, ranging from 500\,nm to 1800\,nm. Applying their results to parameters closest to those of EV Lac, a maximum reduction factor of 5 in RV amplitude between the optical and NIR would be expected. Although this factor should increase for larger spot filling factors or smaller temperature differences, they emphasized that accurate consideration of differences between photospheric and spot spectral features would be essential for more precise estimates. \citet{Barnes_2011_spot-induced-RV} conducted a similar study using the Doppler Tomography of Stars (DOTS) code \citep{Collier_1997_DOTS} and obtained comparable results, finding a reduction factor of 5 between the V and Y bands for small temperature differences (250 K). They also emphasized that in such cases, the ratio of line depths at photospheric and spot temperatures plays a crucial role and could explain trends similar to those we report here. This aspect will be further explored in Sect. \ref{subSect.Contrast2} to provide an explanation for the strong chromaticity observed, following our own flux ratio estimates, detailed in Sect. \ref{subSect.Contrast1}.\\ \indent
In addition to this chromaticity, the activity signal seen in SOPHIE appears to evolve with time on a typical timescale of around a year, as it is highlighted in Fig. \ref{Fig3_SOPHIE_seasons}. The magnetic regions of opposite polarities were described by \citet{Morin_2008_M-B-fields}, and the presence of two spots on the stellar surface was later discussed in \citet{Jeffers_2022_CARMENES-activity}.
More recently, under the assumption of two spots, \citet{Ikuta_2023_spots}  explored their properties using starspot modeling of TESS light curves. They used observations from July 2018 to October 2019, and found a latitude of $25.70^{+0.31}_{-0.39}$ degree for the first spot, and $46.14^{+0.65}_{-0.60}$ degree for the second spot. Our RV SOPHIE data are consistent with an independent evolution of the two spots. The RV-induced amplitude reaches its maximum during the first season for both. Then, the effect of one spot decreases during the second season before recovering its amplitude in the last season. In contrast, the effect of the second spot decreases significantly in the third season.

\subsection{Per-band radial velocity}\label{subSect.Chromaticity2}

\begin{figure}[]
    \centering
    \includegraphics[width=1\linewidth]{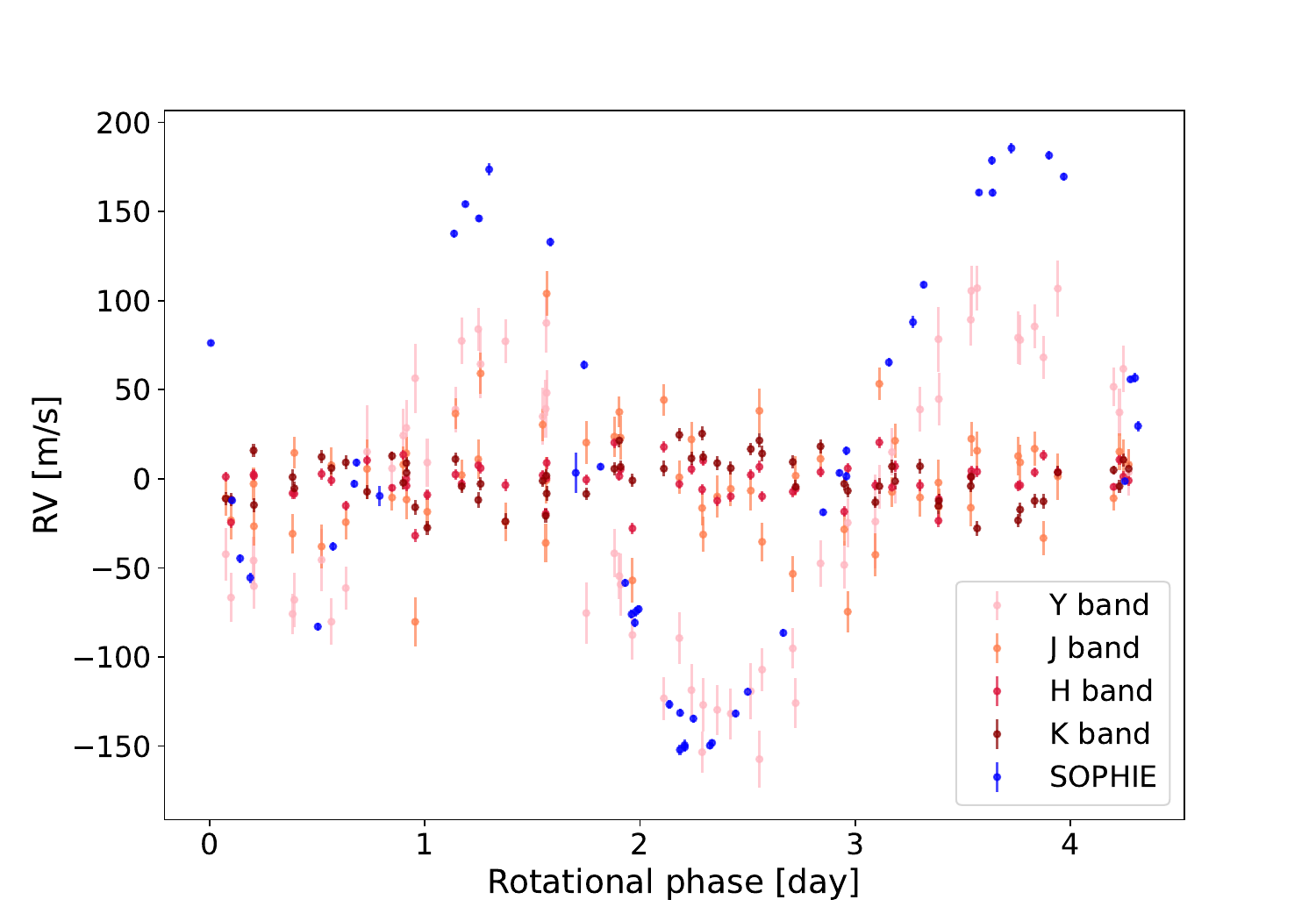}
    \caption{Phase-folded RVs  on $P_{rot}=4.36\,d$, from the different bands of SPIRou, overplotted on SOPHIE full-band RVs.
SOPHIE full band, RMS = \hbox{109.1\,m/s}, mean error = \hbox{3.6\,m/s in blue}.
Y band, RMS = \hbox{75.6\,m/s}, mean error = \hbox{14.3\,m/s in pink}.    \textbf{} J band, RMS = \hbox{30.9\,m/s}, mean error = \hbox{10.5\,m/s in orange}.
    H band, RMS = \hbox{10.5\,m/s}, mean error = \hbox{3.0\,m/s in red}.
    \textbf{} K band, RMS = \hbox{12.7\,m/s}, mean error = \hbox{3.8\,m/s in dark erd}.
    For greater clarity, only the first concomitant season is presented here. Note: these per-band SPIRou data do not follow the same statistical line rejections as the Fig.\ref{Fig1_SOPHIE-SPIRou_RVs} dataset, resulting in higher RMS values.}
    \label{Fig4_bands_RVs}
\end{figure}

To gain a more precise view of the chromaticity of EV Lac RVs, one way is to analyze them for each photometric band. For that, we use LBL per-band pre-computed RVs, corresponding to the different bands of each instruments, namely, U [387, 431\,nm], B [400, 490\,nm], G [450, 500\,nm], V [500, 600\,nm], R [550, 694\,nm] for SOPHIE; and Y [970, 1070\,nm], J [1170, 1330\,nm], H [1490, 1780\,nm], and K [2030, 2370\,nm] for SPIRou.\\  \indent
For the SOPHIE case, each band gives an activity signal that remains similar within the uncertainties, therefore, no chromaticity could be identified at the level of our accuracy (Fig. \ref{B:SOPHIE_bands}). However, in the SPIRou wavelength range, a clear chromatic behavior is detected, as shown in Fig. \ref{Fig4_bands_RVs}: The Y band is particularly sensitive to stellar activity, exhibiting a signal amplitude comparable to that observed in the full band of SOPHIE. Meanwhile, the J, H, and K bands show a progressive decrease in the RV activity signal with increasing wavelength. Notably, the H and K bands do not exhibit a detectable spot-induced activity signal within our accuracy limits of \hbox{3.12\,m/s} and \hbox{4.02\,m/s}, respectively.\\ \indent
These results can be compared with previous studies. In particular, the chromaticity of the rotational signal of EV Lac has been investigated by \citet{Tal-Or_2018_EVLac-CARMENES-chromaticity} and \citet{Jeffers_2022_CARMENES-activity} using the optical arm of CARMENES, which is limited to wavelengths below 960\,nm \citep{Bauer_2020_CARMENES}. Their observations revealed a strong negative chromaticity in the chromatic index (CRX) - a metric quantifying the wavelength dependence of RVs \citep{Zechmeister_2018_CRX} - indicating that the RV amplitude per spectral order decreased with wavelength between 570\,nm and 960\,nm. The absence of this behavior in our SOPHIE data (Fig. \ref{B:SOPHIE_bands}) is likely due to our wavelength coverage ending at 694\,nm. However, \citet{Jeffers_2022_CARMENES-activity} suggested that beyond 950\,nm, the RV scatter would remain constant, contrary to our findings.\\  \indent
To accurately compare our results with theirs and explain the significant decrease in RV amplitude by a factor of 15 between optical and NIR (Fig. \ref{Fig1_SOPHIE-SPIRou_RVs}), we need to analyze the RV chromaticity at higher wavelength resolution, which could lead to a better understanding of its exact physical origin. Therefore, in the next section, we detail our investigation of RV variations in narrower spectral ranges.

\subsection{Per-wavelength bins radial velocity}\label{subSect.Chromaticity3}

\begin{figure}[]
    \centering
    \includegraphics[width=1\linewidth]{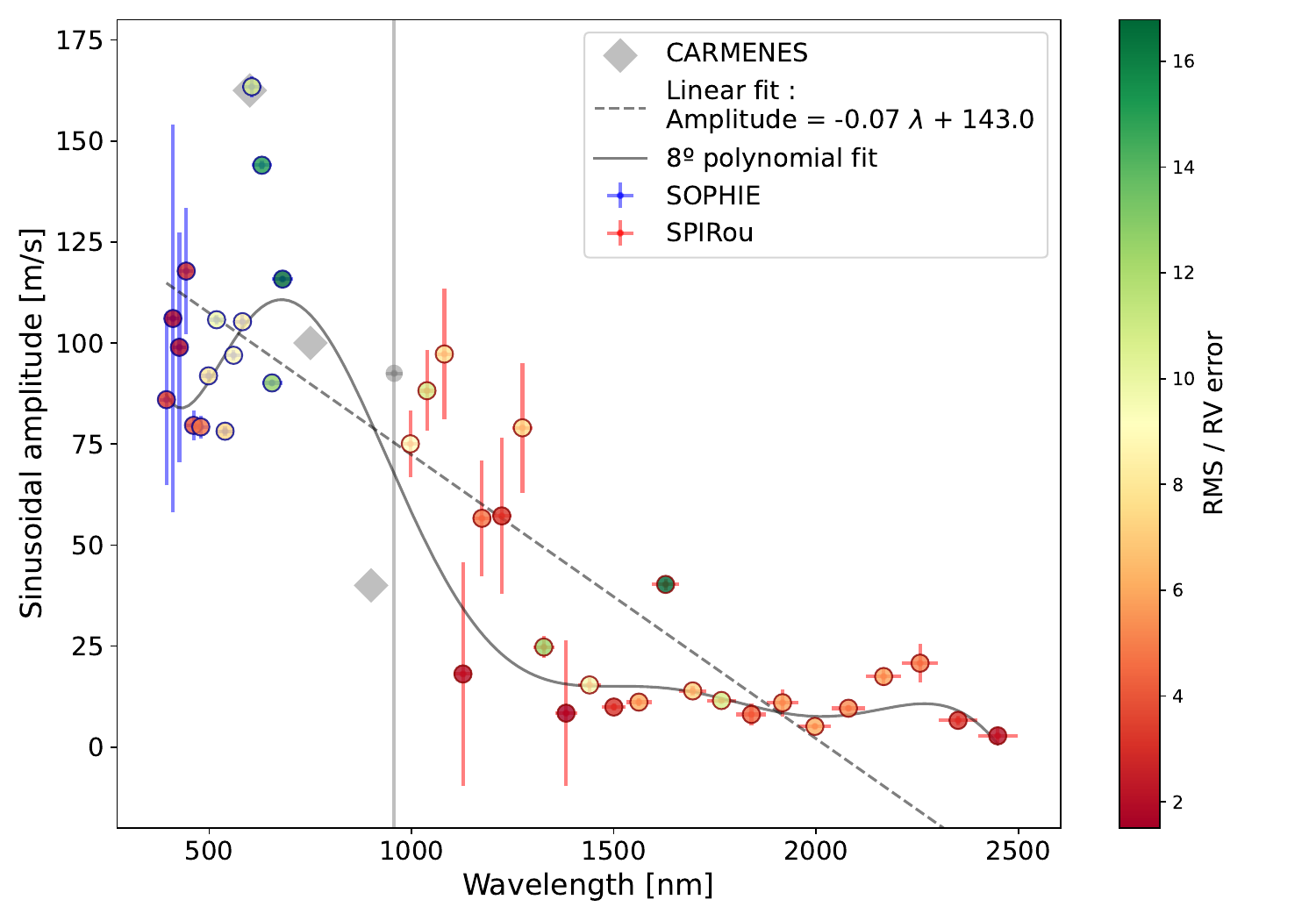}
    \caption{EV Lac's chromaticity plot, showing the amplitude of the activity signal for each wavelength bin, from the SOPHIE to the SPIRou ranges. The dotted line is a linear fit, related to the CRX, while the 8th degree polynomial fit is used as a visual aid to illustrate the trend, which can be compared to the predictions of \citet{Barnes_2011_spot-induced-RV}. The colour bar shows the RMS over the mean RV error. The first bin of the SPIRou dataset is shown in grey and has been removed from the fit due to its very high amplitude uncertainty and very low RMS/RV error. The detail of the sinusoïdal fit for each point can be found in Appendix \ref{Appendix_C:sin_fits}. The three diamond-shaped grey points extracted from the linear fit in Figure 3 of \citet{Jeffers_2022_CARMENES-activity} are included here simply for comparison purposes, as they were not observed at the same epochs.}
    \label{Fig5_chromatic_slope}
\end{figure}

To probe the chromaticity of the activity signal across multiple wavelength bins, we need the right balance between a good sampling in wavelength, while keeping enough spectral information, which means defining a sufficiently wide range of wavelengths for each bin. This work is already done by the LBL pipeline, which returns 15 bins for SOPHIE and 24 bins for SPIRou, evenly distributed in $log(\lambda)$. Each bin provides a RV time series, from which the amplitude of the activity signal is estimated by fitting a sinusoidal function to the phase-folded RV data, with the period fixed at half the rotation period (the sinusoïdal fit for each point can be found in Appendix \ref{Appendix_C:sin_fits}). A sinusoidal fit is used rather than an RMS estimate, as the signal is coherent and periodic, making the fit more physically meaningful than a dispersion-based measure. The significance of these amplitude estimates is indicated by a colorbar representing the ratio of the RMS to the mean RV error for each time series. This ratio helps assess whether the sinusoidal signal is likely to be dominated by photon noise due to a limited number of lines, or whether it reflects a significant activity signal above the noise. The results are shown in Fig. \ref{Fig5_chromatic_slope}.

The conclusions of the previous section are confirmed here, with a clear chromatic slope and a Y band that is as sensitive to activity as the visible range. It is particularly insightful to compare these results with the predictions of \citet{Barnes_2011_spot-induced-RV} (Fig.2 of their paper), who obtained a similar theoretical trend by accounting for the relative line depths between the photosphere and the spot. Building on this, the next section will focus on the effect of line contrast on individual spectral features to further investigate the chromaticity of our measurements.

\section{Toward a contrast-related origin of EV Lac's chromaticity} \label{Sect.Contrast}

The effect of a spot on the RV results from the combined changes in the shape and depth of the spectral lines. In this section, we analyze the sources of these variations to determine how to select the lines based on their impact on the RV.

\subsection{Continuum contrast}\label{subSect.Contrast1}

\begin{figure}[]
    \centering
    \includegraphics[width=1\linewidth]{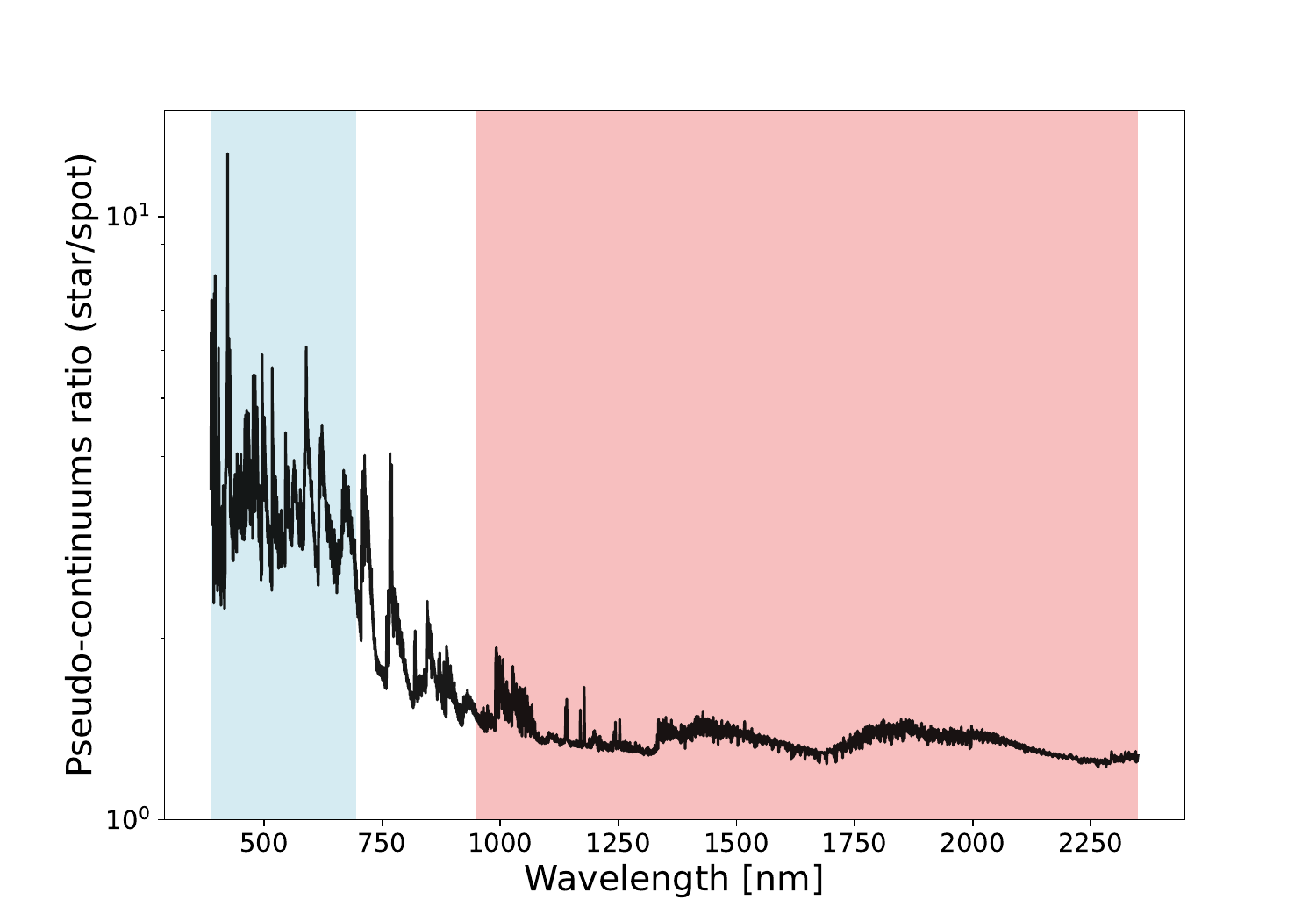}
    \caption{Continuum contrast estimated for EV Lac, using the pseudo-continuums of a PHOENIX photospheric model spectrum at 3300\,K and a spot model spectrum at 3000\,K. \textbf{} SOPHIE wavelength range in blue. SPIRou wavelength range in red.}
    \label{Fig6_Contrast_continuum}
\end{figure}

When considering the impact of stellar spots on RV, one of the strongest effect probably comes from the flux contrast between the spots and the photosphere \citep[e.g.,][]{Reiners_2010_spots}. In the case of EV Lac, the photospheric temperature is estimated at 3340\,K (see Table \ref{Table_Stellar_params}), while we estimate the spots to have a temperature around 3000\,K, as discussed at the end of Section \ref{subSect.EVLac2}. To estimate the resulting contrast, we use PHOENIX model spectra \citep{Husser_2013_PHOENIX} at 3300\,K and 3000\,K for the photosphere and the spots, respectively. The value of \textit{log g} might differ between these regions due to increased magnetic pressure in active features, which reduces gas pressure \citep{Solanki_2003_Starspots, Bruno_2022_starspots} and mimics a stellar model with lower surface gravity. However, for this exploratory study, we assume the same value for both spectra, as investigating this effect is beyond the scope of this paper. The pseudo-continuum is estimated using a 3\,nm rolling average, smoothing out fine spectral lines while preserving broader molecular bands. The photospheric continuum is then divided by the spot continuum. The resulting ratio, defined as the continuum contrast, is shown in Fig. \ref{Fig6_Contrast_continuum}.

While we can see an overall agreement between this continuum contrast and the chromaticity plot of Fig. \ref{Fig5_chromatic_slope}, there are some notable differences, particularly in the Y band, where the amplitude of the RVs is similar to the optical (Fig.\ref{Fig5_chromatic_slope}), for a much lower contrast (Fig. \ref{Fig6_Contrast_continuum}). This indicates that the continuum contrast is not sufficient to fully account for the chromatic behavior of EV Lac's RV measurements. This unexpected  behaviour of chromaticity may arise from other physical sources (e.g., the Zeeman effect, explored in Section \ref{Sect.Zeeman}) or from an incomplete assessment of the contrast effect. We further investigate this latter possibility in the following section.

\subsection{Per-line contrast}\label{subSect.Contrast2}

To push  our definition of the contrast even further, we chose to fully exploit the power of LBL RVs, and try to define in the same way a "per-line contrast." This will enable us to associate a sensitivity to the contrast effect for each line defined by LBL, and select them accordingly. The mathematical framework used to create this new definition is developed in the following subsection, before being applied to the EV Lac data.\\  \indent

\begin{figure*}[]
    \centering
    \includegraphics[width=1\linewidth]{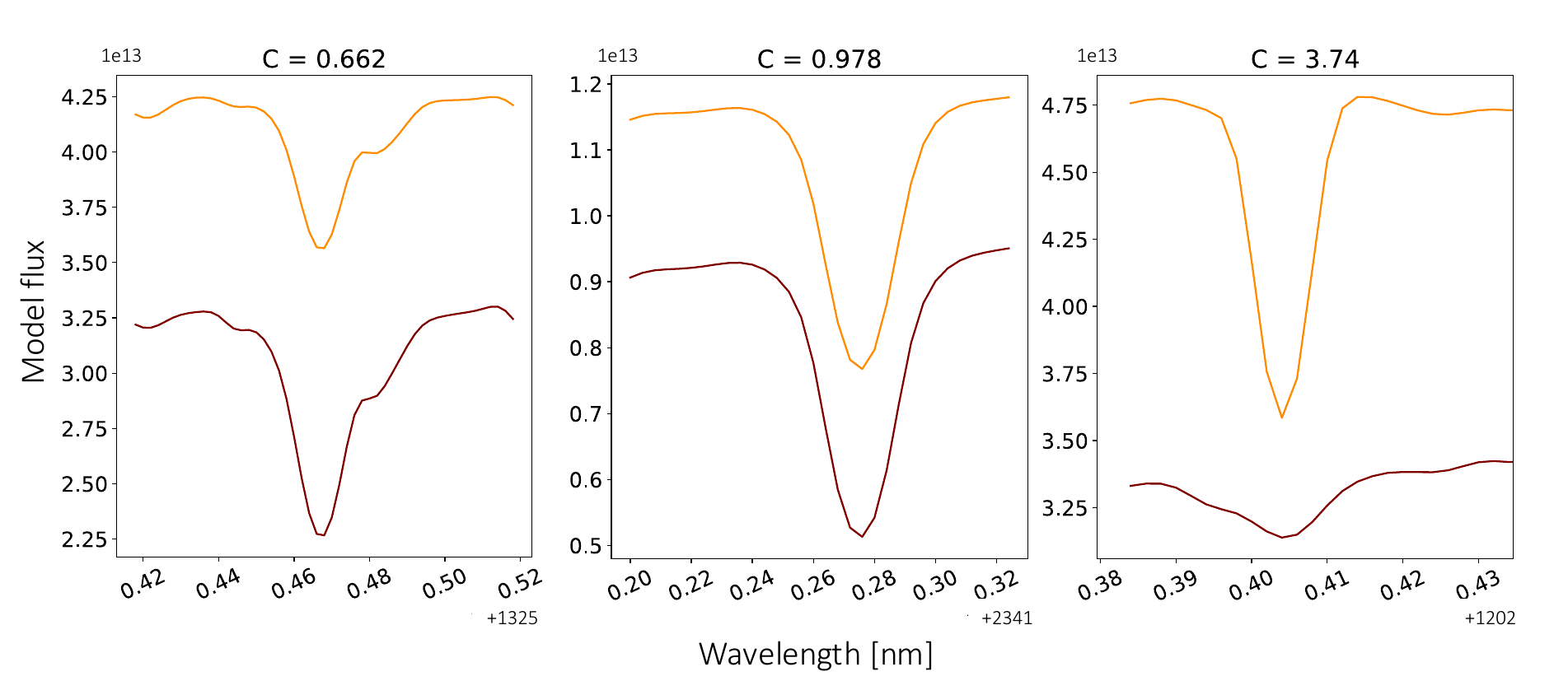}
    \caption{Example lines from PHOENIX models, for the three contrast cases (from left to right: $C<1$, $C\approx1$, $C>1$). \textbf{} Orange lines represent photosphere lines, at 3300\,K. Brown lines represent the spot lines, at 3000\,K.}
    \label{Fig7_example_lines}
\end{figure*}

\subsubsection{Mathematical framework}

The question to answer here is not only how the spot spectrum differs from that of the photosphere, but mostly how this difference impacts the RV computation. Furthermore, this difference needs to be expressed at the line level. For this purpose, we define the observed line profile $L^{\text{obs}}$ as the disk-integrated contribution of a local line profile $l(\lambda, \Delta\lambda, z)$\footnote{The function $l(\lambda, \Delta\lambda, z)$ is defined along a coordinate system where $z$ is aligned with the stellar rotation axis, and $\Delta\lambda$ denotes the position along the axis perpendicular to the rotation axis, as this position is directly related to the rotational Doppler shift of the spectral line.} expressed as

\begin{align}
    L^{\text{obs}} &= \iint l(\lambda, \Delta\lambda,z) \, d\Delta\lambda \, dz. \label{Eq.IV-Lobs}
\end{align}

In an active region, the local physical conditions (e.g.,  magnetic field strength or temperature) differ from those in quiet regions. As a result, the local spectral line profile originating from an active region, $l_A(\lambda, \Delta\lambda, z)$, differs from that of a quiet region, $l_Q(\lambda, \Delta\lambda, z)$. We therefore locally distinguish between quiet and active line profiles. For simplicity, we hereafter refer to $l_A(\lambda, \Delta\lambda, z)$ and $l_Q(\lambda, \Delta\lambda, z)$ simply as $l_A$ and $l_Q$.
Thus, for a quiet star "Q" hosting an active region "A," the observed mean line profile can be expressed as\\

\begin{align}
    L^{{obs}} &= \iint_Q l_Q \, d\Delta\lambda \, dz + \iint_A l_A \, d\Delta\lambda \, dz,\\
    &= \iint_{\mathrm{star}} l_Q \, d\Delta\lambda \, dz + \iint_A \left( l_A - l_Q \right) \, d\Delta\lambda \, dz.
\end{align}

It can be noted that for a given spectral line, if the quantity $\Delta l = l_A - l_Q$ is zero, there will be no distortion in the observed mean line profile. In that case, the RV associated with this line will remain unchanged during the passage of the active region. For a spot that is cooler than the surrounding quiet photosphere, it can be assumed that, for a given line, the temperature difference between the two regions leads, to first order, to a difference in absolute line depth $D$. This can be expressed relative to the continuum level $C$ as a relative depth $D^{\text{rel}}$, so $\Delta l$ becomes

\begin{align}
    \Delta l &= C_A \times D^{\mathrm{rel}}_A \, f(\lambda, \Delta\lambda,z) - C_Q \times D^{\mathrm{rel}}_Q f(\lambda, \Delta\lambda,z),\\
    &= f(\lambda, \Delta\lambda,z) \left( C_A \times D^{\mathrm{rel}}_A - C_Q \times D^{\mathrm{rel}}_Q \right),
\end{align}

where $f(\lambda, \Delta\lambda,z)$ represents the shape of the line independently of its depth and continuum. Thus, the lines that are theoretically "insensitive" to the active region will satisfy the condition:

\begin{align}
    C_A \times D^{\mathrm{rel}}_A - C_Q \times D^{\mathrm{rel}}_Q &= 0\\
    \Leftrightarrow \quad \frac{C_A \times D^{\mathrm{rel}}_A}{C_Q \times D^{\mathrm{rel}}_Q} &= 1 
\end{align}

In the end, this shows that all lines with a given couple (relative depth, continuum) will have the same influence on the overall observed line profile. Thus, the spot-to-photosphere contrast of a line, which we  simply refer to as the contrast for the remainder of this paper, is defined here as

\begin{equation}
    \hfill Contrast = \frac{D^\mathrm{rel}_\mathrm{photosphere}}{D^\mathrm{rel}_\mathrm{spot}} \times \frac{C_\mathrm{photosphere}}{C_\mathrm{spot}}. \hfill
    \label{eq_contrast}
\end{equation}

Three specific cases can be distinguished:

\begin{itemize}
    \item $Contrast > 1$: The photosphere contributes more to the overall line than the spot. This line will produce a "normal" RV signal during the transit of a spot. We refer to this regime as the "normal contrast" regime.
    \item $Contrast = 1$: The spot contributes equally to the overall line as the photosphere. This line will produce no RV variation during the transit of a spot. We refer to this as the "neutral contrast" regime.
    \item $Contrast < 1$: The spot contributes more to the overall line than the photosphere. This line will produce a "reverse" RV signal during the transit of a spot (i.e., a RV shift that is first blueshifted, then redshifted). We refer to this as the "anti-contrast" regime, which is different from the contrast generated by hot regions.\footnote{This formalism could be applied to hotter regions, such as faculae or plages, by computing the contrast using a spectrum hotter than the photosphere. However, this is beyond the scope of the current paper, and may not be relevant for EV Lac, where the spots dominate the activity.}
\end{itemize}

Ultimately, we demonstrate that the depth of the line is a crucial parameter to consider when accurately accounting for the impact of a spot on RV measurements. This was previously highlighted by \citet{Reiners_activity_2010}, who theoretically examined the case of lines with a contrast of 1, where the presence of a spot would be undetectable in RVs. However, their study did not explore the case of anti-contrasted lines, for which spots induce RV signals anti-correlated with those typically associated with starspots. From a physical perspective, such a scenario could arise, for instance, if certain lines that are absent from the photosphere only form at cooler temperatures, within the spot. It is important to clarify that this "anti-contrast" does not refer to a hot spot, but rather to a cold spot that contains more spectral information than the surrounding photosphere, for some given lines.\\ \indent

\subsubsection{Radial velocities in the different contrast regimes}

\begin{figure*}[]
    \centering
    \includegraphics[width=1\linewidth]{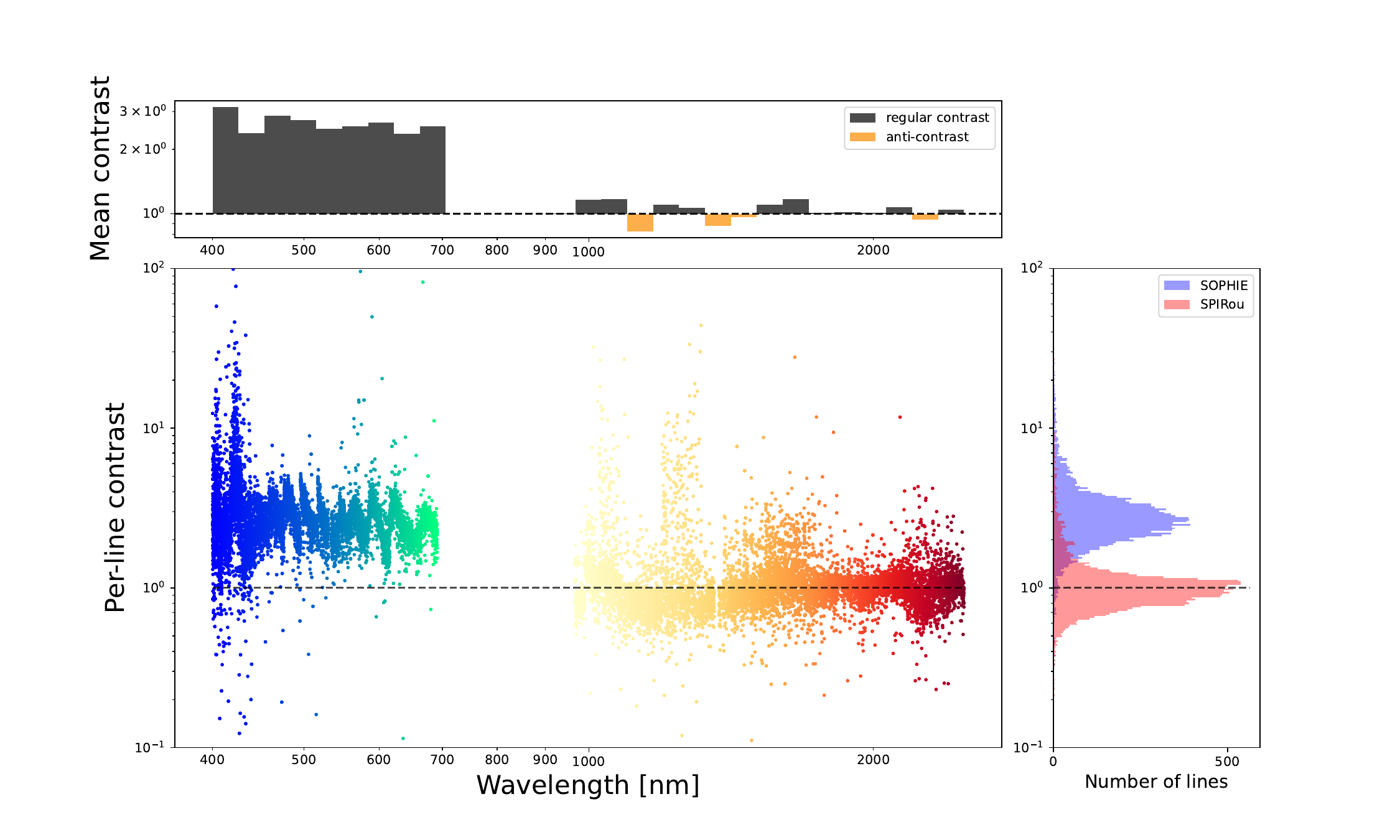}
    \caption{Contrast computed for each line. The neutral contrast C=1 is highlighted by the dotted lines. \textbf{Middle:} Contrast estimation across the SOPHIE to SPIRou wavelength range. \textbf{Right:} Histogram of contrast values for SOPHIE and SPIRou. \textbf{Top:} Contrast averaged in wavelength bins for SOPHIE and SPIRou, computed using a geometric mean to capture the proportional nature of the per-line contrast (e.g., the geometric mean $\bar{C}_{\mathrm{geo}} = \left( \prod_{i=1}^{N} C_i \right)^{1/N}$ equals 1 for opposite contrasts, such as $C=0.5$ and $C=2$). The bar heights are proportional to the expected RV amplitude, while their color encodes the phase of the signal.}
    \label{Fig8_per-line-contrast}
\end{figure*}

To calculate the per-line contrast in the case of EV Lac, we first create our LBL line list skeleton following the methodology described in Section \ref{Sect.Methodology}. For each of these wavelength intervals (or line) defined by LBL, we look at the corresponding intervals in the two PHOENIX models used previously: for the spot as well as the photosphere model, the continuum is estimated by taking the mean of the beginning and the end of the interval (theoretically corresponding to the extremity of the wings). As the lines can be as small as a few PHOENIX wavelength points, we won't average on more points. The relative depth is defined as the difference between the line's minimum flux and the continuum, normalized by the continuum. \\  \indent
Computing this contrast from Eq. \ref{eq_contrast} for each line reveals that the three distinct contrast regimes are indeed represented, as illustrated by three random lines in Fig. \ref{Fig7_example_lines}. The wavelength distribution of this per-line contrast in the SOPHIE and SPIRou domain is shown in Fig. \ref{Fig8_per-line-contrast}, where each point is the spot-to-photosphere contrast value of a given line. The neutral contrast (C=1) for which the spot should have no impact on the RVs is highlighted by the dotted line, separating the regular (above) from the anti-contrast (below) regimes. A wealth of information can be extracted from this Fig.\ref{Fig8_per-line-contrast}.\\ \indent
Firstly, we see again the chromatic slope, which can be compared to the one we had when considering the continuum contrast only in Fig. \ref{Fig6_Contrast_continuum}. However, many more details are present, especially some lines in the Y band that appear to be highly contrasted, explaining the high RV signal in this region. Also, the higher amplitude seen around 1600\,nm in Fig. \ref{Fig5_chromatic_slope} does correspond to a region of high contrasted lines. Even more interesting is the distribution of anti-contrast lines: while they are almost non-existent in the SOPHIE domain, they account for around half of the lines in the SPIRou domain. This result indicates that, in the present case, about half of the lines influence the RV in the opposite way to the other half.
This is visible both in the right-hand histogram, where SPIRou contrasts are symmetrically distributed around C=1, and in the upper plot: the relatively small bar heights for SPIRou, compared to SOPHIE, suggest a lower RV amplitude, while the presence of anti-contrasted bars indicates that, at certain wavelengths, the activity signal should be reversed relative to the visible range. Overall, this histogram suggests that the H and K bands are particularly insensitive to EV Lac's spots, which is consistent with the results shown in Fig.~\ref{Fig4_bands_RVs}.

\begin{figure}[]
    \centering
    \includegraphics[width=0.9\linewidth]{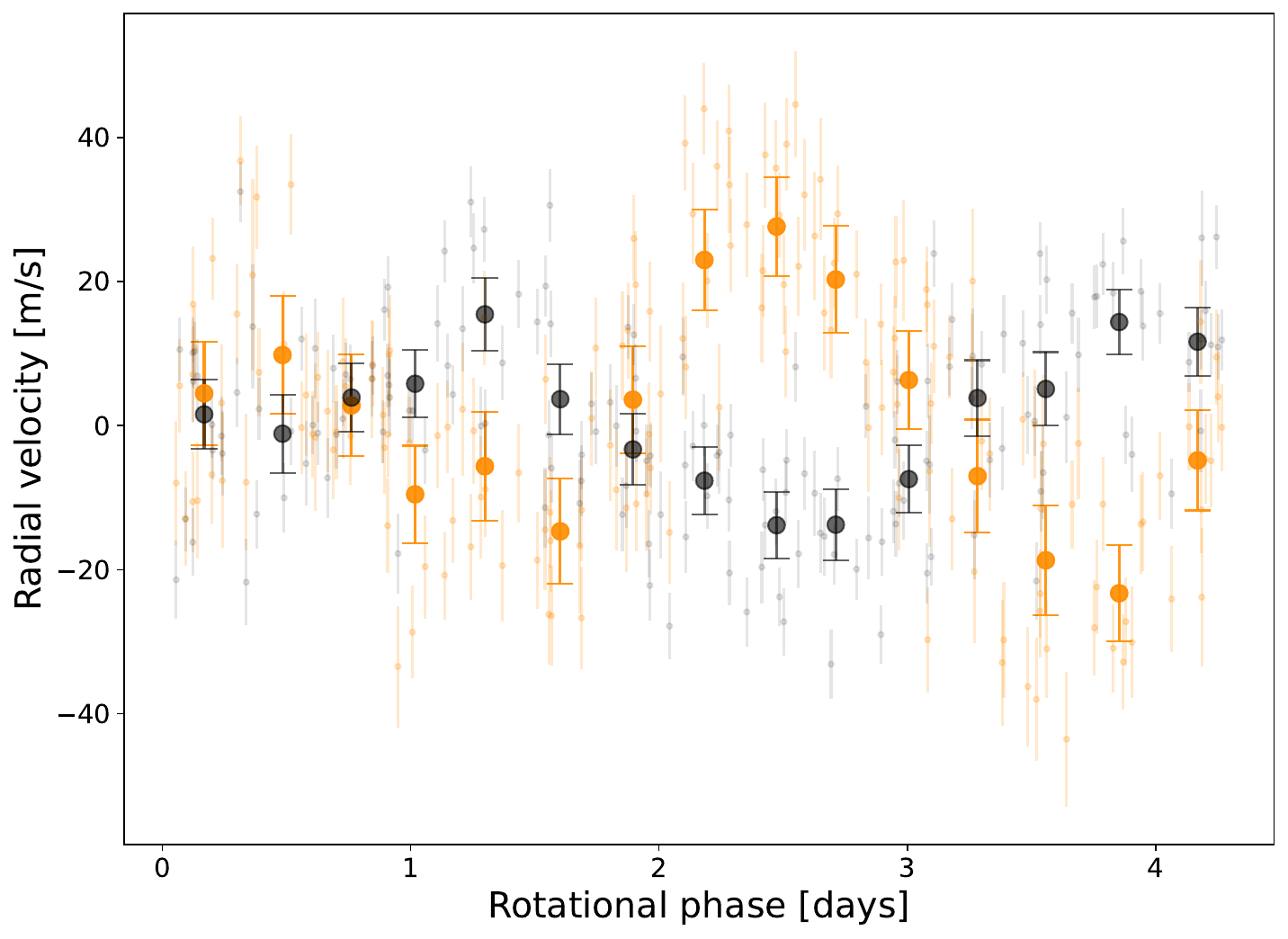}
    \caption{RV time series phase-folded on the rotation period (4.36 days), comparing contrasted and anti-contrasted lines. \textbf{} Time series using 7180 lines with C>1.05, which behave similarly to a "regular spot" that hides a part of the photosphere, shown in black (RMS = \hbox{13.7\,m/s}). \textbf{} Time series using 9464 lines with C<0.95, which behave similarly to an "anti-contrasted spot",  adding more spectral information than the photosphere, shown in yellow (RMS = \hbox{19.1\,m/s)}.}
    \label{Fig9_anti_usual-contrast}
\end{figure}

To validate our understanding of this phenomenon, we separated the SPIRou lines into two contrast types and analyzed their RV signals. Since lines around C = 1 should be insensitive to the spot, we excluded them to better highlight the two anti-correlated contrasts, with an arbitrary threshold of $\pm 0.05$ chosen from the histogram in Fig.\ref{Fig8_per-line-contrast}. Using our line selection method, we computed an anti-contrast time series with 9,464 lines ($C<0.95$) and another with 7,180 lines ($C>1.05$). The results are shown in Fig. \ref{Fig9_anti_usual-contrast}, where a striking anti-correlation between the two time series is observed, providing observational evidence that anti-contrast may play a role as significant as, if not more important than, regular contrast. In that specific case, the two time series have a RMS around \hbox{14\,m/s} for the regular contrast and \hbox{19\,m/s} for the anti-contrast, which yield for a factor 5 and 7 of amplitude diminution between SOPHIE and these new SPIRou signals. This indicates that the very low activity signal seen when computing the RVs using all infrared lines (Fig.\ref{Fig1_SOPHIE-SPIRou_RVs}) is due to contrast and anti-contrast being two anti-correlated effects of similar amplitude, effectively canceling each other out.\\  \indent
In conclusion, our new definition of line contrast empirically demonstrates our ability to disentangle two families of line-dependent RVs, whose strong anti-correlation provides a clear and necessary explanation for the pronounced chromaticity observed between the two instruments. It is interesting to note that similar anti-correlated RVs had already been observed in the past on $\alpha$ Cen b by \citet{Cretignier_2020_LBL_pRVs_II} and \citet{AlMoula_2022_LBL_pRVs_III}, who rightly thought that they were linked to a certain temperature sensitivity, albeit originating from convection inhibition.\\  \indent

Concerning the lines in the wavelength domain of SOPHIE, there are too few anti-contrasted lines to generate an anti-contrast time series. This means that an activity filtering approach based on line selections, as tested in Sect. \ref{Sect.Correction}, would probably not be relevant in the optical domain.

\section{Probing the Zeeman effect}\label{Sect.Zeeman}

Given the strong magnetic field of EV Lac (Table \ref{Table_Stellar_params}), we investigated the potential role of the Zeeman effect in the chromaticity of the RV signal. As demonstrated by \citet{Reiners_2013_Zeeman} and \citet{Hebrard_2014_Zeeman}, the strength of this effect increases with the square of the wavelength, implying a strong positive chromaticity (i.e., a greater impact in the near-infrared), unlike the contrast effect.\\ \indent
Using the VALD (Vienna Atomic Line Database; \citealt{Piskunov_1995_VALD, Kupka_2000_VALD, Ryabchikova_2015_VALD, Pakhomov_2019_VALD}), we identified lines with known high effective Landé factors ($g_\mathrm{eff}$), which quantify the sensitivity of spectral lines to Zeeman broadening. To disentangle this effect from temperature contrast, we further excluded lines exhibiting excessively strong (regular or anti) contrast. We therefore computed RVs using a subset of lines with $g_\mathrm{eff} > 1.2$ and contrast values within the arbitrary interval $C \in [0.75, 1.333]$\footnote{This interval corresponds to a symmetric selection in contrast space, defined as $[1/x, x]$. The chosen thresholds aim to reduce the impact of strongly contrasted lines, while retaining enough spectral information.}.\\ \indent
This analysis revealed no clear evidence for an increase in the amplitude of the activity signal when applying the above-mentioned line selections. In the case of SOPHIE, most lines exhibit strong contrast (see Fig.~\ref{Fig8_per-line-contrast}), leaving only 35 usable lines after selection, which is too few to obtain meaningful constraints, given the resulting large uncertainties. For SPIRou, both the dispersion and the associated uncertainties increased to comparable levels, rendering any definitive interpretation inconclusive. Moreover, because effective Landé factors are primarily available for atomic lines, and often missing for the molecular transitions that dominate the NIR, this selection likely underestimates the true impact of the Zeeman effect in that spectral domain.\\ \indent
Nevertheless, to limit the potential impact of the Zeeman effect as much as current data allow, we conservatively exclude lines with Landé factors $g_\mathrm{eff} > 1.2$ in the remainder of this work.

\section{Toward a correction of spot-induced activity}\label{Sect.Correction}

Once we probed both the temperature contrast and the Zeeman effect of EV Lac, we  tried to use this knowledge to filter out the RV activity signal. Concerning the Zeeman effect, we simply mitigated it by removing lines with an effective Landé factor above a given threshold. We chose to remove lines with $g_\mathrm{eff}>1.2$, as this provides a good balance between eliminating the most affected lines and preserving the majority of lines and, thus, the signal.\\ \indent
Regarding the contrast effect, the most logical approach would be to select only lines with a neutral contrast ($C\approx1$). However, this would not only result in the removal of a large number of lines but would also rely on the assumption that the calibration of $C$ is perfect (a point we discuss in Sect. \ref{subSect.Correction2}).

\subsection{Impact of the spot temperature} \label{subSect.Correction2}

Selecting lines with $C\approx1$ should isolate lines that are "neutral" in terms of temperature contrast, meaning they are not distorted when spots pass by. However, this requires an accurate calibration of $C$, meaning the correct determination of photosphere and spot temperatures, as well as the ability of PHOENIX models to accurately reproduce their emitted flux. While these models have slight inaccuracies, they still allow for a statistical classification of lines into two families — regular contrast and anti-contrast. But a key source of error remains the poor accuracy in determining the spot temperature as noted in Section \ref{subSect.EVLac2}. Both of these effects may slightly offset the neutral contrast value.\\  \indent
We might take this issue to our advantage though. If lines with $C\approx1$ still exhibit an activity signal, it may indicate that the temperature adopted for the spot is incorrect. In this case, an iterative adjustment of the spot temperature to minimize the activity signal on these lines could serve as a new method to determine it. For that purpose, we computed the line contrast using temperatures within the estimated range, from 2800\,K \citep{herbst_2021_starspots} to 3200\,K \citep{Ikuta_2023_spots} by steps of 100\,K. For each set of contrast values associated to a given temperature, we independently compute the RVs for the three contrast regimes (regular, neutral and anti-contrast). We arbitrarily chose $C \in [1.05, 4]$ for the regular contrast regime, $C \in [1/1.05, 1.05]$ for the neutral contrast regime and $C \in [1/4, 1/1.05]$ for the anti-contrast regime. Then, for each resulting time series, we estimated the activity signal amplitude by fitting a sinusoid at $P_{\text{rot}}/2$. Additionally, we excluded lines with $g_\mathrm{eff}>1.2$, as discussed above. We expected  the spot temperature closest to reality would yield a null RV signal for $C\approx1$ and two symmetric RV signals for the lines with regular contrast and anti-contrast. 

\begin{figure}[]
    \centering
    \includegraphics[width=1\linewidth]{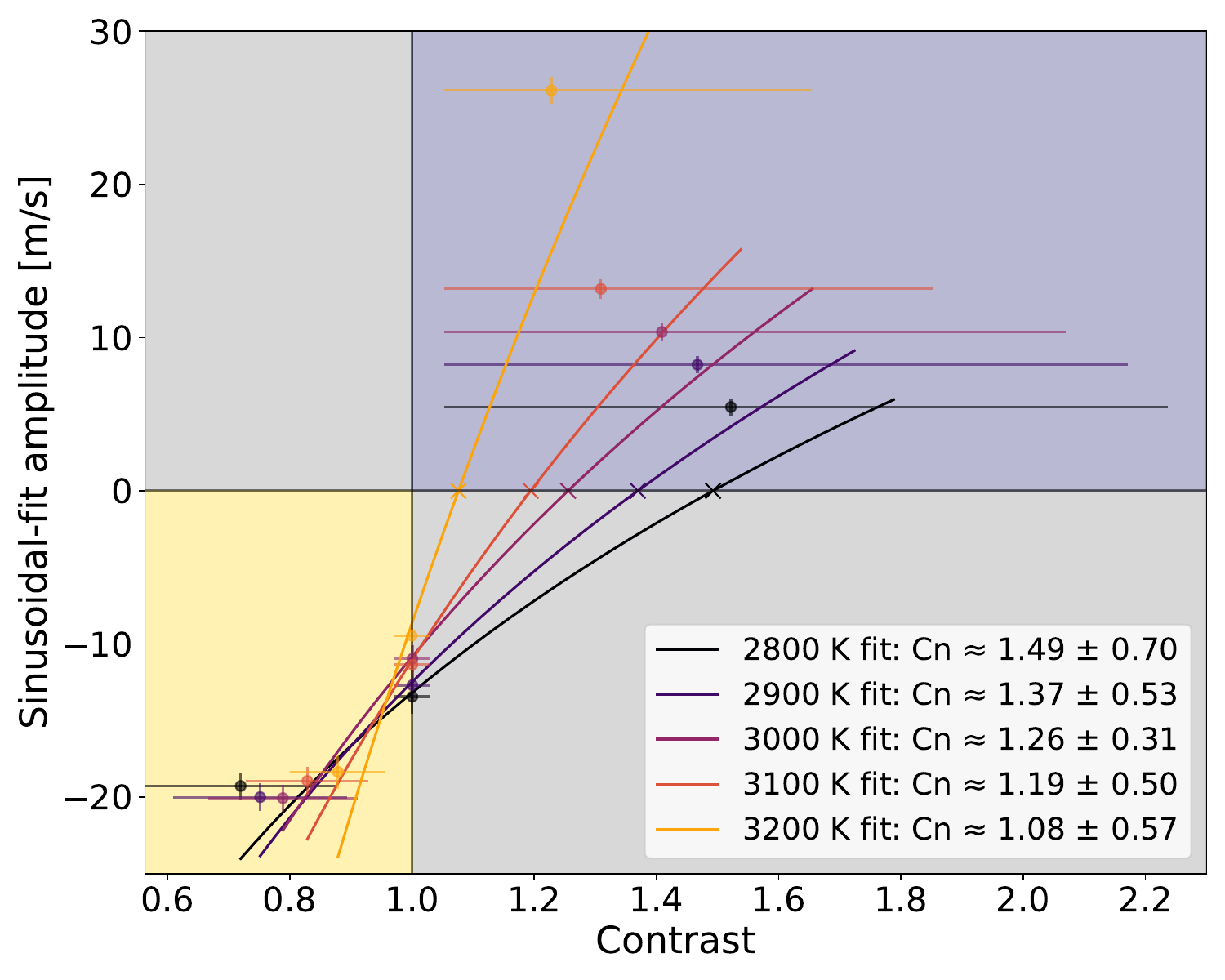}
    \caption{Evolution of the sinusoidal fit amplitude for RV time series computed using spectral lines in different contrast regimes, for a photosphere at 3300K and different spot temperatures. The sign of the amplitude gives the phase of the signal (Negative $\Rightarrow$ anti-contrast phase, Positive $\Rightarrow$ regular contrast phase). Contrast values are computed as the mean of per-line contrasts within each regime, with standard deviations as uncertainties. For each temperature, a linear fit in log(contrast) space is used to determine the effective neutral contrast $C_n$, namely, the contrast at which the signal amplitude is expected to vanish. Region corresponding to time series exhibiting an anti-contrast phase, shown in yellow. Region corresponding to time series  with a regular-contrast phase, shown in blue. {\bf } Areas indicating regions where the time series should not appear if the spectra from the spot and the photosphere were perfectly modeled, shown in gray.}
    \label{Fig11_spot_temperatures}
\end{figure}

Figure \ref{Fig11_spot_temperatures} presents the results of these computations. The sign of the amplitude indicates the phase of the sinusoidal fit, with a negative sign corresponding to an anti-contrast regime and vice versa. Each contrast value corresponds to the average of the distribution of the per-line contrasts within a given regime, with the uncertainty corresponding to its standard deviation. For each spot temperature, we performed a linear fit of the activity-induced RV amplitude in the "log(contrast)" versus "amplitude" space. This allows us to estimate the value of C for which the amplitude approaches zero. Ideally, if the modeling of the emergent spectra of the photosphere and the spot were perfect, this would occur at $C = 1$. In practice, spectral lines with $C$ values close to 1 tend to exhibit an anti-contrast signal, and this effect becomes even stronger at lower spot temperatures (i.e., when the temperature difference $\Delta T$ between the spot and the photosphere is larger). This suggests that the spectral line modeling is not entirely accurate, and that the model with the highest spot temperature, while not perfect, provides the best approximation.\\  \indent
These results suggest that a small temperature contrast is preferred in the case of EV Lac, with a spot temperature slightly above 3200K. This aligns with the conclusions of \citet{Ikuta_2023_spots} presented in Section \ref{subSect.EVLac2}. However, several limitations may introduce biases in this conclusion, as discussed in the final section.

\subsection{Balancing contrasts to mitigate activity}\label{subSect.Correction1}

\begin{figure*}[]
    \centering
    \includegraphics[width=1\linewidth]{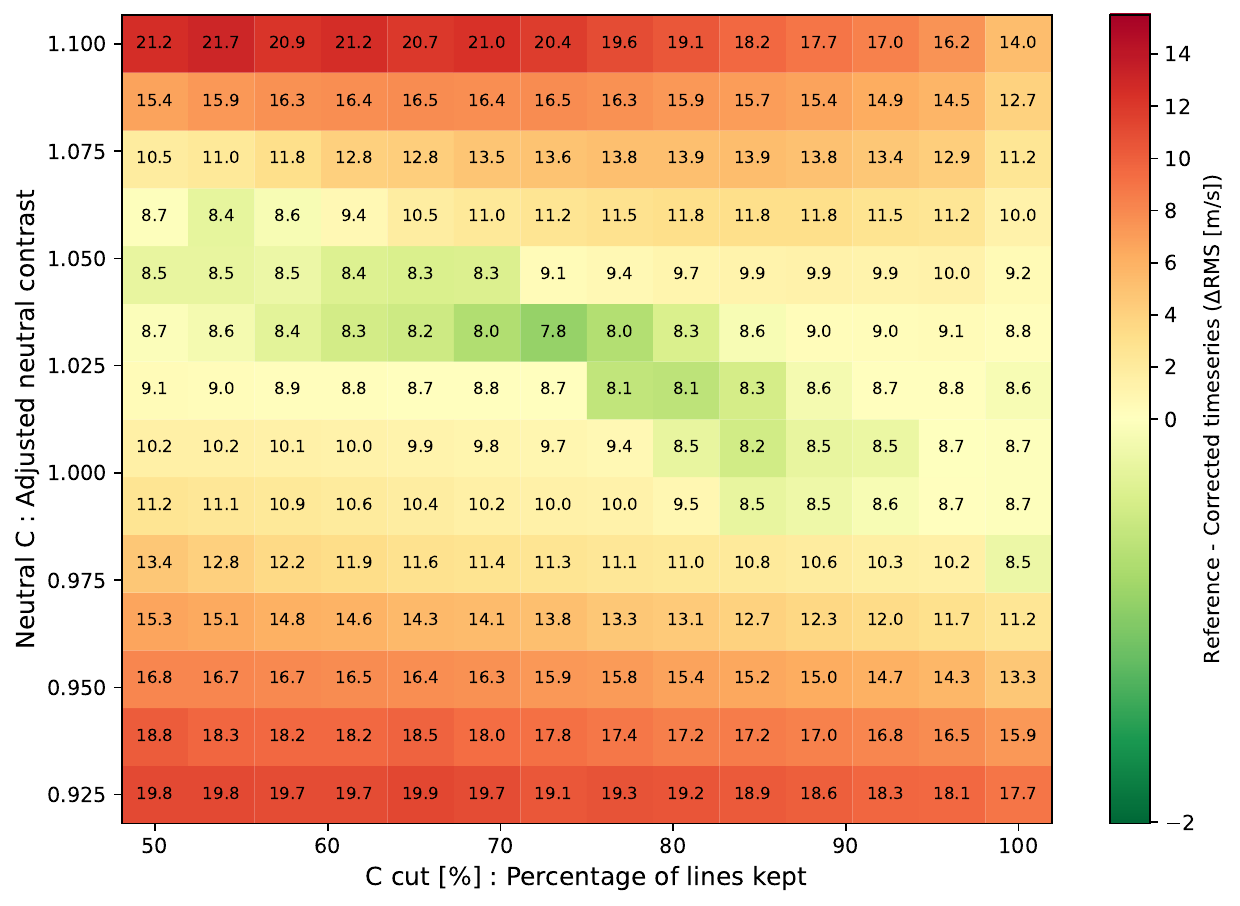}
    \caption{Comparison between the reference and the corrected time series for a photosphere at 3300K and a spot at 3200K. The RMS of the corrected time series are overprinted (in m/s), while the colorbar shows the difference of RMS between the reference (RMS = 8.70\;m/s) and the corrected time series. RMS reduced by the process, in green. RMS increased by the process, in red.}
    \label{Fig10_RMS_MAP}
\end{figure*}

We recalculated the contrast of all lines using our spot temperature determination at 3200\,K, and propose a method to filter out the activity signal, based on the knowledge developed in this study. As outlined in Section \ref{Sect.Contrast}, no filtering was performed on the SOPHIE dataset, as nearly all lines are sensitive in the same way to spot-photosphere temperature contrast, making the selection of activity-insensitive lines ineffective. However, in the case of SPIRou dataset, a sufficient number of lines exist in all three contrast regimes. Consequently, the following analysis is conducted exclusively on the SPIRou dataset (seasons 1 and 2).\\  \indent
The most straightforward filtering strategy is to select only neutral-contrast lines. However, as discussed in Section~\ref{subSect.Correction2}, even lines with contrast values close to unity still exhibit a residual activity signal likely due to limitations in spectral modeling (Fig. \ref{Fig11_spot_temperatures}), making this approach unreliable. Moreover, such a selection would have excluded too many lines to preserve an acceptable level of statistical uncertainty.\\  \indent
However, another approach can be undertaken in this context. By chance, on this star the distribution of contrasts for lines in the SPIRou range appears to be fairly balanced between regular and anti-contrast (see Fig. \ref{Fig8_per-line-contrast}), leading to a partial cancellation of the two effects. However, slight imbalances may still exist between the line lists of the two regimes. In fact, if we manage to perfectly balance the two effects, we might be able to effectively cancel the contrast-induced activity signal. This requires not only accounting for the number of lines in each contrast category but also considering their contribution to the RV computation, which depends on the spectral information derived from the inverse of the LBL per-line error.\\  \indent
To separate regular and anti-contrast lines, the theoretical selection threshold should be set at C = 1. However, as discussed in the previous section, even with perfect spectral modeling, this neutral contrast assumption holds only if the temperatures used for the photosphere and spot models are accurate. Yet, even with a spot temperature of 3200\,K, Fig. \ref{Fig11_spot_temperatures} showed that the neutral contrast remains slightly above 1. Also, since PHOENIX models are defined in 100\,K increments, there is room for minor offsets. Therefore, we define this neutral contrast threshold as a variable, referred to as $C_{\text{neutral}}$ hereafter.\\  \indent
Furthermore, lines with extreme contrast values may have a disproportionately strong impact on the overall RVs, warranting their removal. This introduces a second variable, denoted as $C_{\rm cut}$\,[\%], representing the percentage of lines with the most extreme contrast values that are excluded from the analysis.\\  \indent
Considering a spot temperature of 3200 K, the classification of lines into regular and anti-contrast categories now depends on these two parameters. To evaluate their impact on RV measurements, we compute histograms of the number of lines weighted by their contrast value divided by their mean error, for both $C > 1$ and $C < 1$ line lists\footnote{For the anti-contrast regime ($C < 1$), the weight is given by the inverse of the contrast value, divided by the mean error.}. The total weight of each histogram represents the contribution of the respective line list to the RVs. We then iteratively remove lines with the most extreme C values (i.e., those farthest from 1) from the more heavily weighted histogram until the two histograms have equal weight. This ensures that the regular and anti-contrast line lists exert an equal influence on the overall RVs. Finally, we merge all lines from the two balanced histograms to create a spot-induced activity corrected line list, from which we compute the corresponding time series. The process can be repeated for different pairs of parameters $(C_{\text{neutral}},\, C_{\text{cut}})$.\\  \indent
To assess the effectiveness of this correction for a given pair of parameter, we compare the RMS of the resulting time series to a reference case that includes all LBL lines. However, this reference might contain other physical effects that could bias our assessment of contrast-related activity correction. The only additional effect considered in this paper being the Zeeman effect, we once again ensure that Zeeman-sensitive lines are removed beforehand by excluding those with an effective Land\'e factor above 1.2 (see Section \ref{Sect.Zeeman}). This yields a reference time series with an RMS of \hbox{8.70\,m/s} for a mean error of \hbox{3.50\,m/s}.\\  \indent
The comparison between the \hbox{8.70\,m/s} RMS of this reference time series and the RMS of the corrected time series provides an estimate of the correction's effectiveness. Since the corrected time series depends on the chosen parameters, we repeat the histogram balancing process multiple times to map the $(C_{\text{neutral}}, \,C_{\text{cut}})$ parameter space.\\ \indent
The results of this process, illustrated in Fig.\ref{Fig10_RMS_MAP}, are promising. While most of the parameter space shows that line selection worsens the final RMS (compared to the time series including all lines except those with a high effective Land\'e factor), there is a narrow region where our line selection effectively reduces activity-induced variations.\\ \indent
Regarding $C_{\text{cut}}$, we observe that rejecting a higher percentage of lines generally increases the RMS due to larger uncertainties, although this trend is not always pronounced. However, keeping all lines never significantly reduces the RMS, regardless of the chosen $C_{\text{neutral}}$ value.\\  \indent
On the other hand, $C_{\text{neutral}}$ behaves as expected, with the RMS increasing on either side of its optimal value, which appears to be around $C_{\text{neutral}} = 1.03$ in this case. This slightly elevated value is consistent with expectations based on Fig. \ref{Fig11_spot_temperatures}.\\  \indent
The best result is obtained by removing approximately 25\% of the lines and setting $C_{\text{neutral}} = 1.03$. In this configuration, the correction reduces the RMS by nearly \hbox{1\,m/s} within a very narrow parameter space, demonstrating the efficacy of this method. However, it remains highly sensitive to an accurate determination of the neutral contrast. Given that we previously reduced the RMS by \hbox{0.22\,m/s} by removing Zeeman-sensitive lines, the overall activity mitigation process through line selection appears promising. It should be emphasized, however, that EV Lac may represent a special case, in which the two contrast regimes are naturally well balanced. This method could therefore prove even more effective when applied to other active M dwarfs.\\  \indent
Ultimately, this method requires precise determination of both the spot temperature and the sensitivity of each spectral line to temperature contrasts, two aspects that can be further improved and are discussed in Section ~\ref{Sect.Conclusion}.

\section{Summary and perspectives}\label{Sect.Conclusion}

In this study, we combine visible and NIR spectroscopic measurements of the young and magnetically active M3.5V dwarf EV Lac. Even though a strong activity signal was detected in the visible at half the stellar rotation period, no such signal was initially found in the NIR.\\  \indent
The absence of an activity signal in the NIR is surprising in light of previous studies, which predict that the Zeeman effect should dominate at these wavelengths for highly magnetic stars. To assess its impact, we used the Landé factors of atomic lines from SOPHIE to SPIRou to identify those most sensitive to the Zeeman effect. This approach enabled us to isolate the Zeeman-induced activity signal in SPIRou. However, the presence of numerous molecular lines with unknown Landé factors in the NIR complicates a precise evaluation of this effect. Although filtering out lines with high Landé factors slightly reduces the overall RV dispersion, the improvement remains limited, while introducing a small increase in measurement uncertainty.\\  \indent
In addition, we report an unexpectedly strong negative chromaticity, which we attribute to temperature contrasts between two oppositely rotating spots. However, continuum flux variations between models at the photospheric and spot temperatures alone cannot explain the observed chromaticity levels. To resolve this discrepancy, we introduced a new contrast definition that accounts for spectral line depths, revealing an "anti-contrast" behavior in a subset of NIR lines. This effect arises from the presence of spots slightly cooler than the photosphere and is so pronounced that it affects nearly half of the NIR lines, leading to a self-cancellation between anti-contrasted and regular contrasted time series. Although this has been observed here for EV Lac, this effect may also be present in a wide range of targets, as it appears, to the first order, to depend primarily on the temperatures of the photosphere and the spots.\\  \indent
This newly identified property provides a potential means of constraining starspot temperatures. Specifically, the contrast value at which we observed a phase reversal in the activity signal depends on the temperature difference between the photosphere and the spots. By comparing this phase reversal for different PHOENIX models, we estimated the spot temperature of EV Lac to be slightly above 3200\,K.\\  \indent
Leveraging this effect, we developed a filtering method to adjust the two anti-correlated time series and try to achieve near-complete cancellation of spot-induced RV variations. Despite this effect being naturally well compensated in EV Lac, our technique still reduced the overall RV dispersion by approximately \hbox{1\,m/s}. However, as the method requires case-by-case application, a future study on different stars would be necessary to assess its effectiveness. For now, it remains a proof of concept with potential improvements to explore. \\  \indent
One limitation of our approach, for instance, is that we adopted a simplified two-temperature model, whereas spots are known to be complex structures often associated with faculae and plages, as well as umbra and penumbra regions with temperature gradients. While modeling spots as a cooler photosphere provides a first-order correction, a more physically realistic spot spectrum or a multi-temperature approach could improve the accuracy. Additionally, EV Lac hosts two dominant spots, which were treated identically in this analysis. Accounting for potential differences in their properties may refine the method further.\\  \indent
Finally, this newly identified class of anti-contrasted lines raises broader questions regarding its occurrence. Namely, we ought to consider how  this effect varies with stellar temperature, spectral type, wavelength range, or metallicity. Addressing these aspects could help identify target populations where the self-cancellation mechanism naturally mitigates activity-induced RV variations.\\  \indent
Ultimately, our study highlights the critical role of multi-wavelength observations in characterizing and mitigating stellar activity in M dwarfs. In this particular case, it has enabled the identification of an anti-contrast effect in starspots. Investigating this phenomenon in future analyses could provide new opportunities for improving stellar activity corrections, thereby enhancing the detection of exoplanets around active M dwarfs.


\begin{acknowledgements}
      This work is based on observations collected with the SOPHIE spectrograph on the 1.93m telescope at the Observatoire de Haute-Provence (OHP) in France, operated by the Institut National des Sciences de l'Univers (INSU) of the Centre National de la Recherche Scientifique (CNRS). We thank the staff of the Observatoire de Haute-Provence for their support at the 1.93 m telescope and on SOPHIE.
      Based on observations obtained with SPIRou, an international project led by the Institut de Recherche en Astrophysique et Planétologie, Toulouse, France, installed at the Canada-France-Hawaii Telescope (CFHT) which is operated from the summit of Maunakea by the National Research Council of Canada, the INSU of the CNRS of France, and the University of Hawaii. The observations at the CFHT were performed with care and respect from the summit of Maunakea which is a significant cultural and historic site. The authors wish to recognize and acknowledge the very significant cultural role and reverence that the summit of MaunaKea has always had within the indigenous Hawaiian community. We are most fortunate to have the opportunity to conduct observations from this mountain.
      This work has made use of the VALD database, operated at Uppsala University, the Institute of Astronomy RAS in Moscow, and the University of Vienna.
      P.L, X.D., A.C., T.F., and X.B., acknowledge funding from the French ANR under contract number  ANR\-24\-CE49\-3397 (ORVET) and in the framework of the Investissement d'Avenir program (ANR-15-IDEX-02), through the funding of the "Origin of Life" project of the Grenoble-Alpes University. 
      S. Bellotti acknowledges funding by the Dutch Research Council (NWO) under the project "Exo-space weather and contemporaneous signatures of star-planet interactions" (with project number OCENW.M.22.215 of the research programme "Open Competition Domain Science- M".
      C. Cadieux, É. Artigau acknowledges that this work is partly supported by the Natural Science and Engineering Research Council of Canada, the Canadian Space Agency and the Trottier Family Foundation through their support of the Trottier Institute for Research on Exoplanets (IREx). This work benefited from support of the Fonds de recherche du Québec – Nature et technologies (FRQNT), through the Center for Research in Astrophysics of Quebec.
      A. Santerne acknowledges support by the "Programme National de Planétologie" (PNP) of CNRS/INSU co-funded by CNES and the "Action Thématique de Physique Stellaire" (ATPS) of CNRS/INSU PN Astro co-funded by CEA and CNES.
      N. Santos acknowledges funding by the European Union (ERC, FIERCE, 101052347). Views and opinions expressed are however those of the author(s) only and do not necessarily reflect those of the European Union or the European Research Council. Neither the European Union nor the granting authority can be held responsible for them. This work was supported by FCT - Fundação para a Ciência e a Tecnologia through national funds by these grants: UIDB/04434/2020 DOI: 10.54499/UIDB/04434/2020, UIDP/04434/2020 DOI: 10.54499/UIDP/04434/2020.
\end{acknowledgements}

\bibliographystyle{aa}
\bibliography{biblio}

\onecolumn
\begin{appendix} 

\section{Periodograms} \label{Appendix_A:periodograms}

Generalized Lomb-Scargle (GLS) periodograms \citep{Lomb_1976_periodogram, Scargle_1982_periodogram, Zechmeister_2009_periodogram}, computed for SOPHIE and SPIRou entire datasets (Fig. \ref{Fig1_SOPHIE-SPIRou_RVs}), using the astropy Python package. 

\begin{figure}[h]
    \centering
    \includegraphics[scale=0.5]{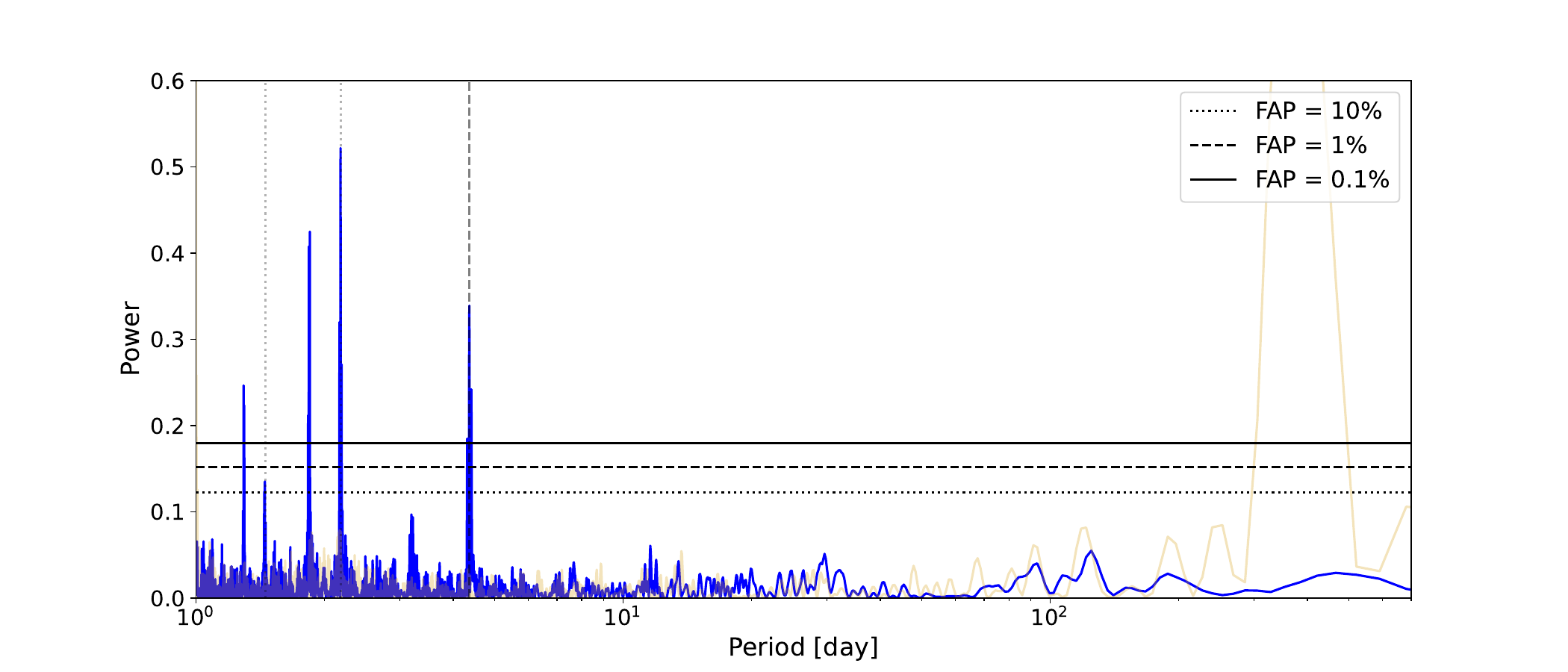}
    \caption{Periodogram of SOPHIE measurements. $P_{rot}$ is highlighted by the grey dashed line, while $P_{rot}/2$ and $P_{rot}/3$ are shown in the light grey dotted lines.}
    \label{A.A:SOPHIE_perio}
\end{figure}

\begin{figure}[h]
    \centering
    \includegraphics[scale=0.5]{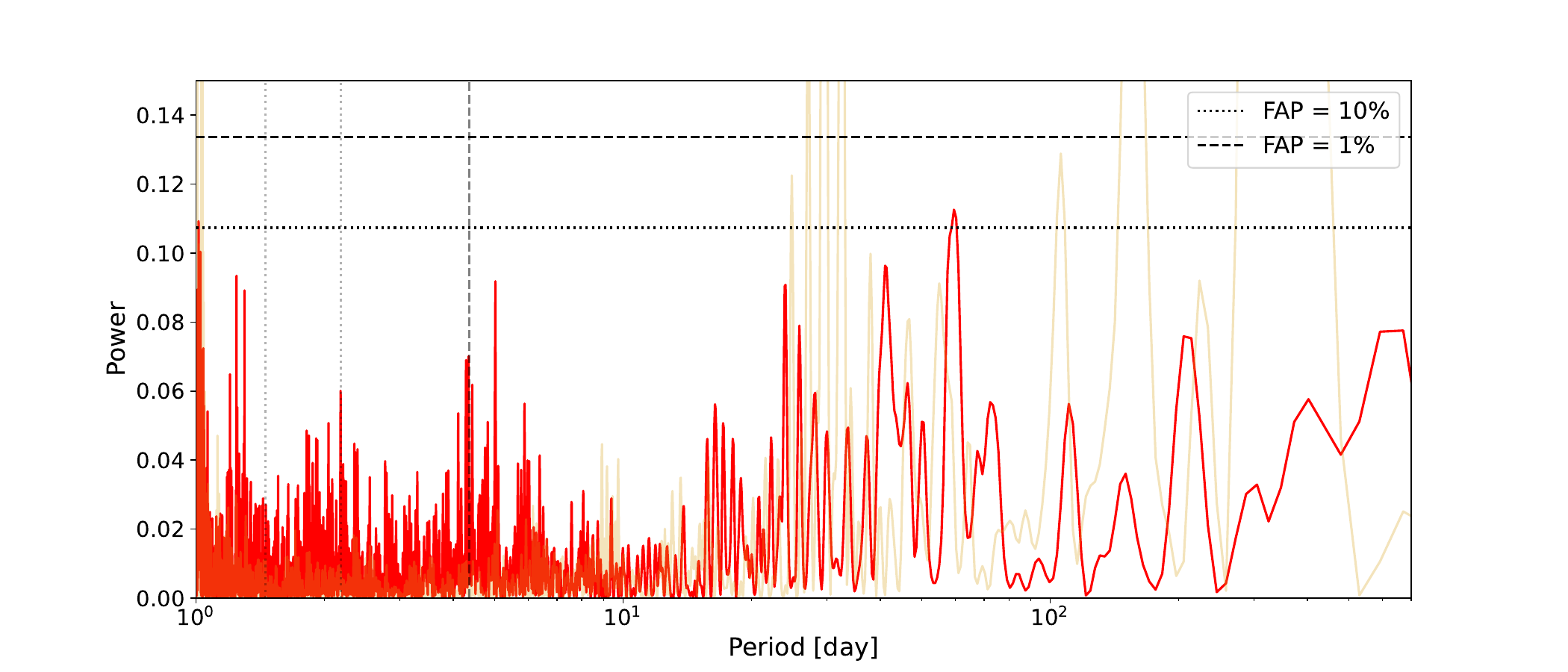}
    \caption{Periodogram of SPIRou measurements. $P_{rot}$ is highlighted by the grey dashed line, while $P_{rot}/2$ and $P_{rot}/3$ are shown in the light grey dotted lines.}
    \label{A.A:SPIRou_perio}
\end{figure}

\newpage

\section{SOPHIE per-band RVs}

As can be seen in the Fig.\ref{B:SOPHIE_bands} below, the chromaticity between the different bands of SOPHIE is not clear, given the uncertainties. The fact that we don't find the negative CRX found on CARMENES by \citet{Tal-Or_2018_EVLac-CARMENES-chromaticity, Jeffers_2022_CARMENES-activity} is probably due to our limited wavelength coverage.

\begin{figure}[h]
    \centering
    \includegraphics[scale=0.6]{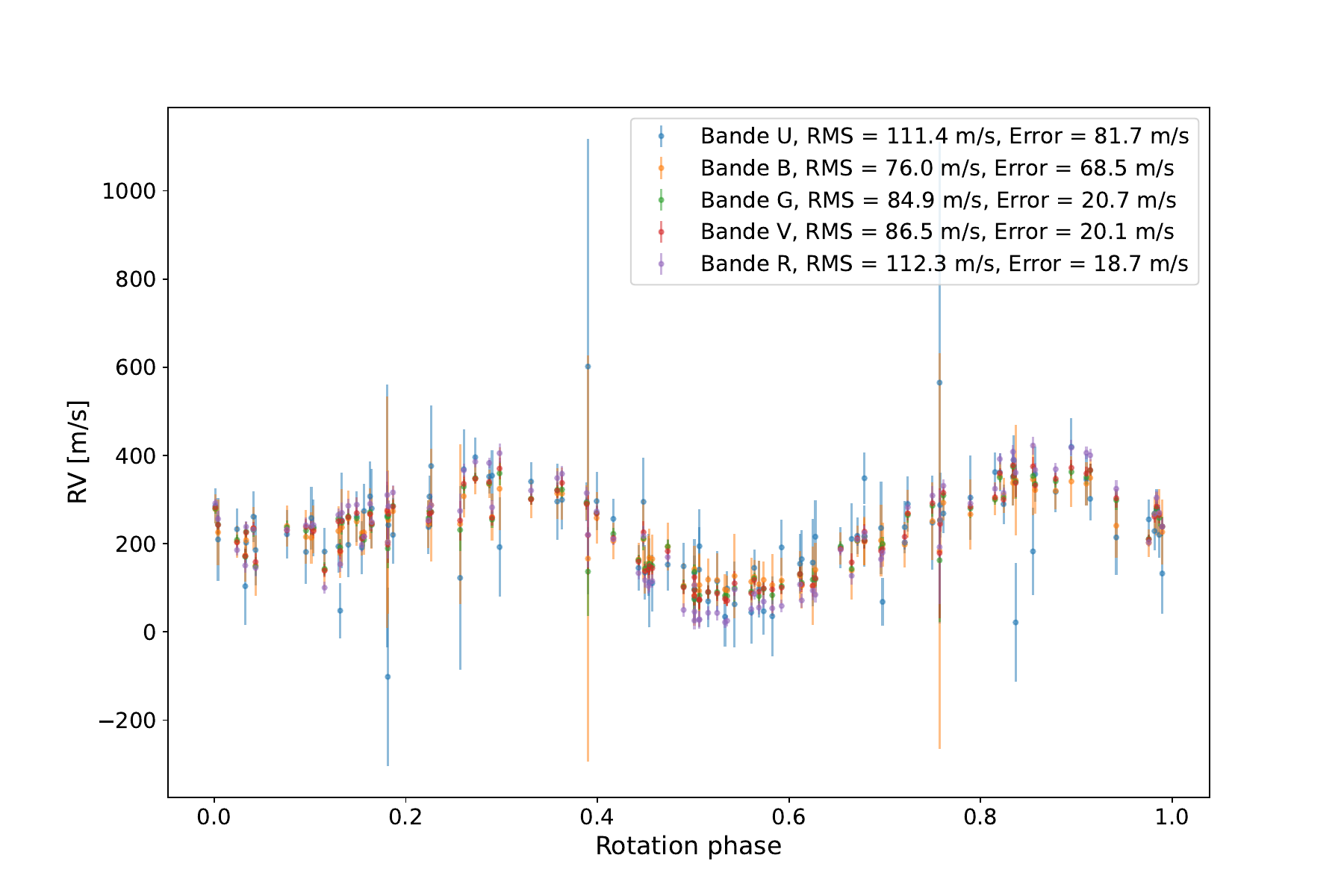}
    \caption{RVs phase-folded on EV Lac rotational period (4.36 days), for the different optical bands of SOPHIE.}
    \label{B:SOPHIE_bands}
\end{figure}

\newpage

\section{Per-wavelength bin RV sinusoïdal fits}\label{Appendix_C:sin_fits}

\begin{figure}[h]
    \centering
    \includegraphics[scale=0.36]{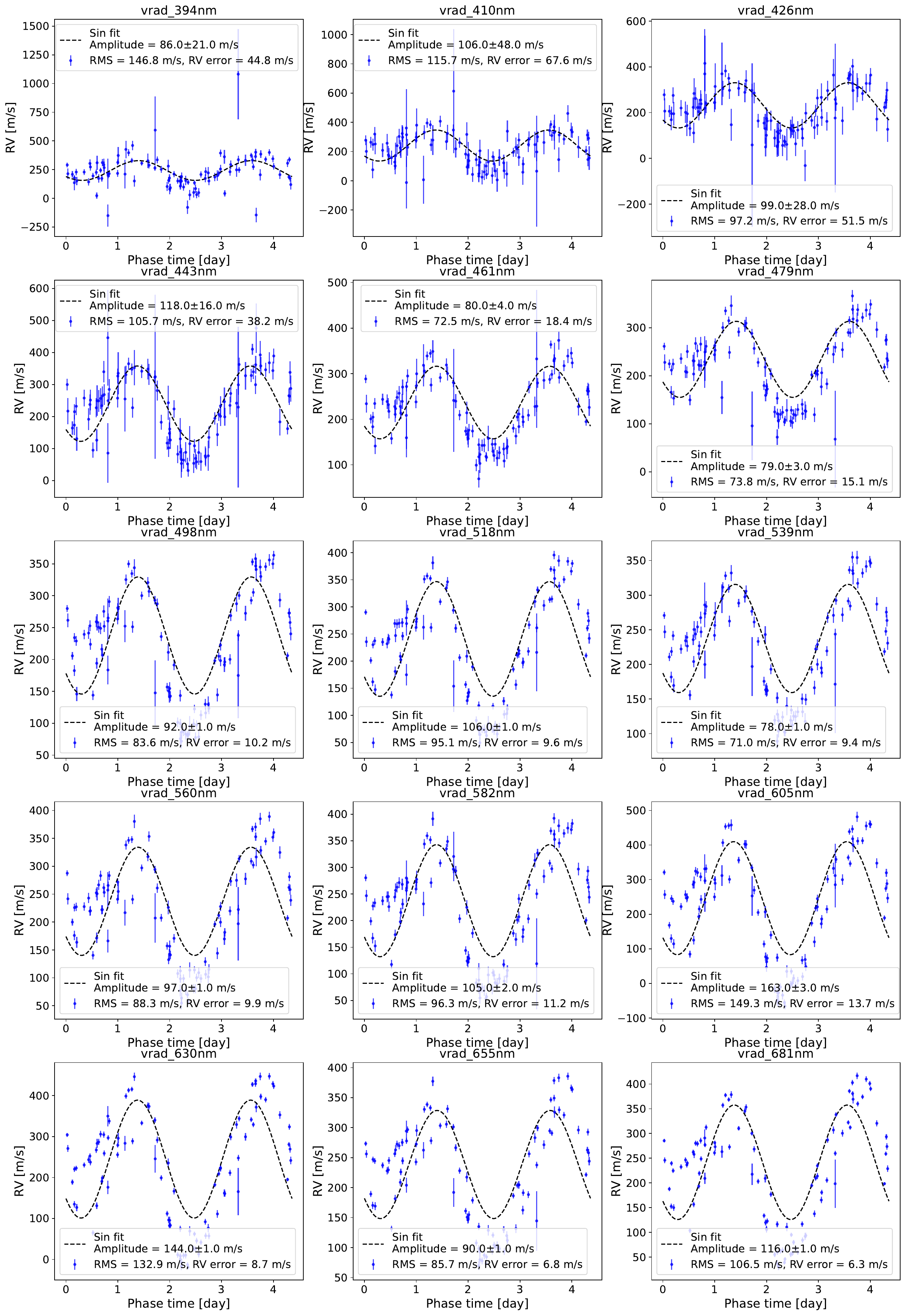}
    \caption{Sinusoïdal fits of the phase-folded SOPHIE RVs for the different wavelength bins considered in Sect. \ref{subSect.Chromaticity3}.}
    \label{App:SOPHIE_sin-fits}
\end{figure}

\begin{figure}[h]
    \centering
    \includegraphics[scale=0.36]{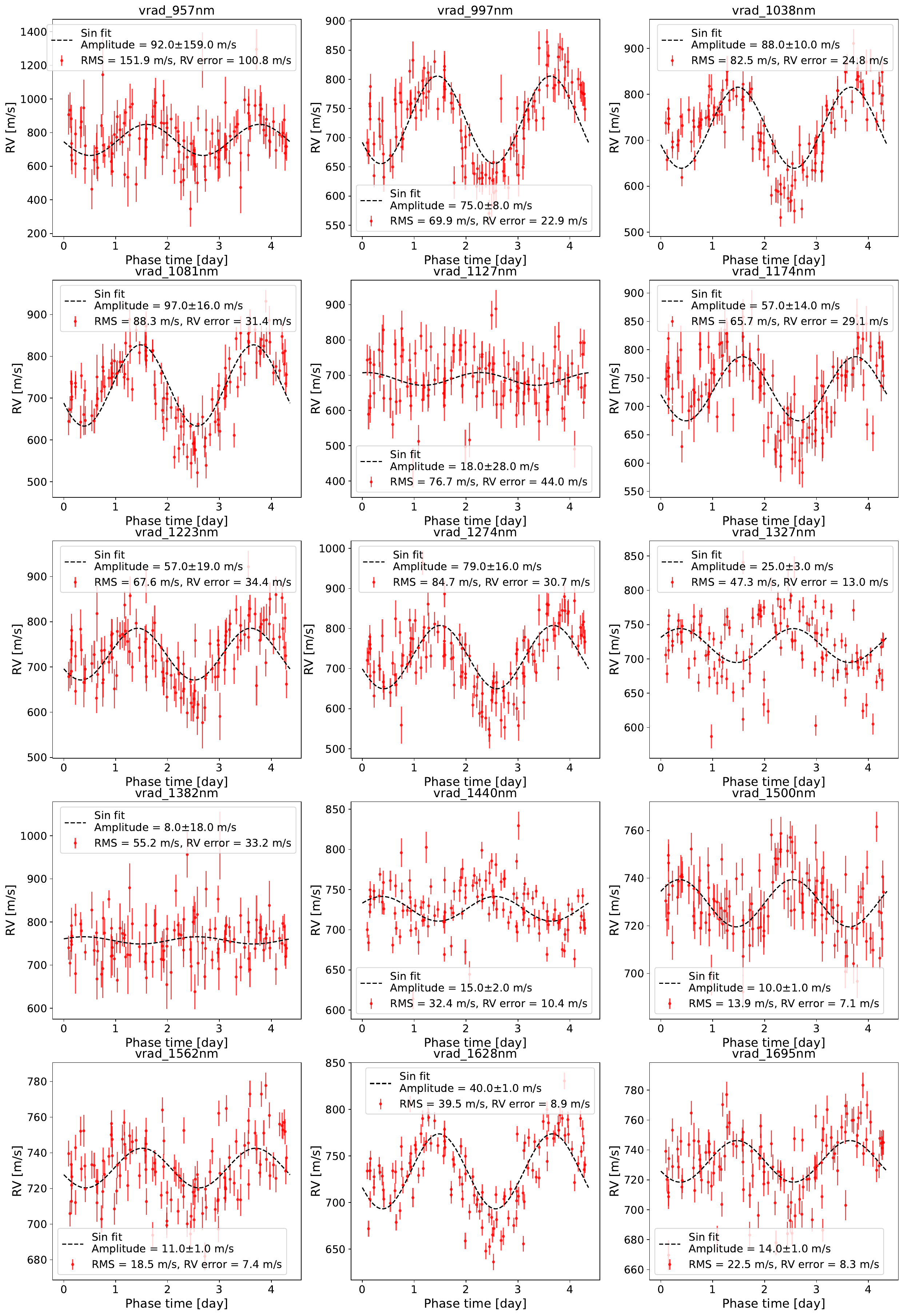}
    \caption{Sinusoïdal fits of the phase-folded SPIRou RVs for the different wavelength bins considered in Sect. \ref{subSect.Chromaticity3}.}
\end{figure}

\begin{figure}[h]
    \centering
    \addtocounter{figure}{-1} 
    \includegraphics[scale=0.4]{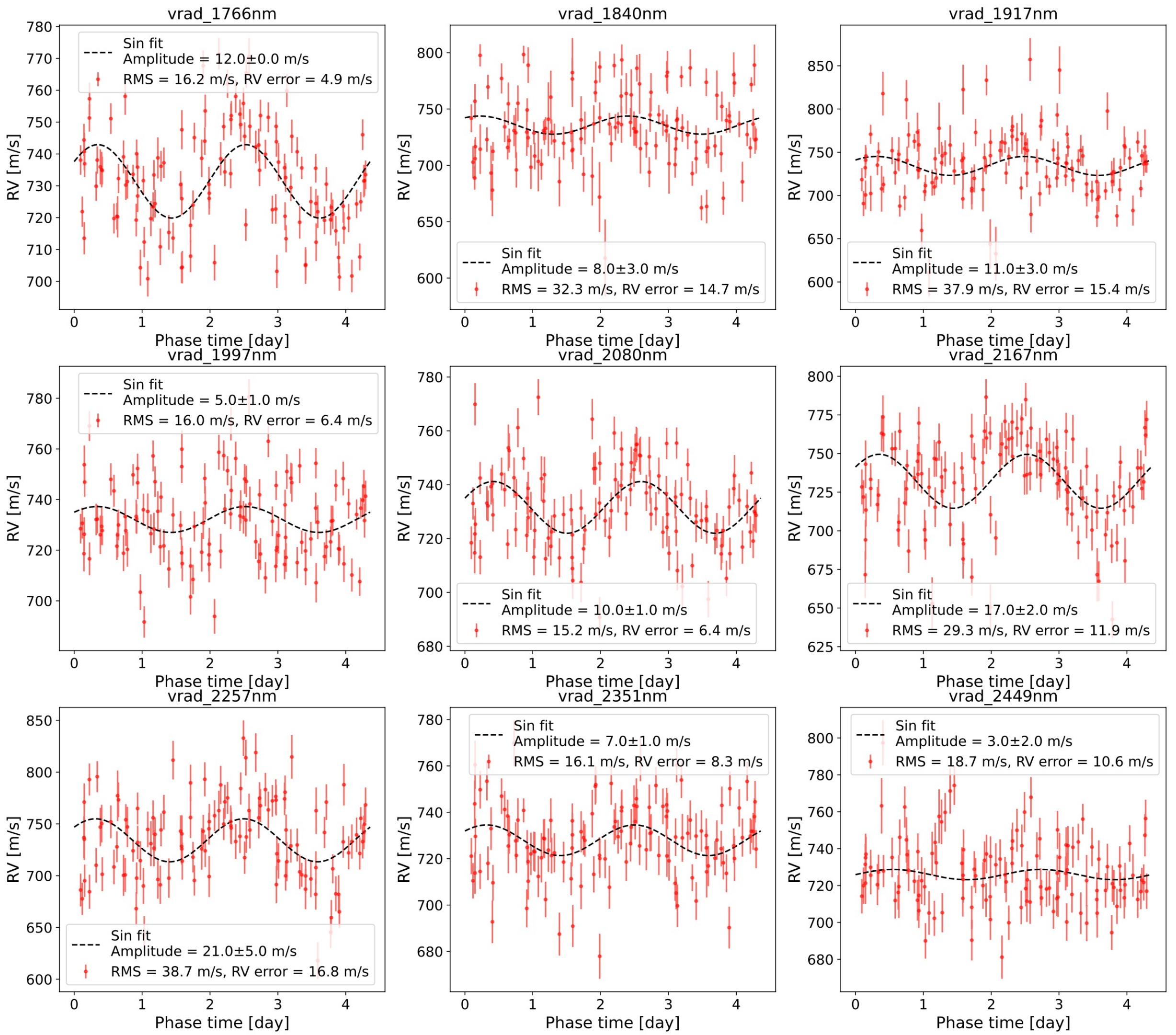}
    \caption{continued.}
    \label{App:SPIRou_sin-fits}
\end{figure}

\end{appendix}

\end{document}